\documentclass[landscape]{aa}
\usepackage{graphicx,psfrag,amsmath,lscape}

\newcommand{\kms}{$\mathrm{\,km\,s}^{-1}$ }
\newcommand{\Vrot}{$\mathrm{V_{rot}}$ }
\newcommand{\Zmax}{$\mathrm{Z_{max}}$ }
\newcommand{\Teff}{$\mathrm{T_{eff}}$ }
\newcommand{\logg}{$\log \mathrm{g}$ }
\newcommand{\FeH}{$\mathrm{[Fe/H]}$ }
\newcommand{\MgFe}{$\mathrm{[Mg/Fe]}$ }
\newcommand{\SiFe}{$\mathrm{[Si/Fe]}$ }
\newcommand{\NiFe}{$\mathrm{[Ni/Fe]}$ }
\newcommand{\alfe}{$[\alpha/\mathrm{Fe}]$ }
\begin{document}
\title{On the correlation of elemental abundances with kinematics among
galactic disk stars.
\thanks{Based on spectra collected with the ELODIE spectrograph at the
1.93-m telescope of the Observatoire de Haute Provence (France)}}
\titlerunning{[Mg/Fe] vs kinematics in the disk}
\author{T.V. Mishenina \inst{1}, C. Soubiran \inst{2},
        V.V. Kovtyukh  \inst{1}, and S.A. Korotin \inst{1}}
\authorrunning{Mishenina et al.}
\offprints{tamar@deneb.odessa.ua}
\institute{
Astronomical Observatory of Odessa
National University and Isaac Newton Institute of Chile, Odessa Branch, Ukraine
\and
Observatoire Aquitain des Sciences de l'Univers, CNRS UMR 5804,
BP 89, 33270 Floirac, France
}
\date{Received \date; accepted}
\abstract{
We have performed the detailed analysis of 174 high-resolution
spectra of FGK dwarfs obtained with the ELODIE echelle spectrograph at
the Observatoire de Haute-Provence.
 Abundances of Fe, Si and
Ni have been determined from equivalent widths under LTE approximation,
whereas
abundances of Mg have been determined under NLTE approximation using
equivalent widths of 4 lines and profiles
of 5 lines. Spatial velocities with an accuracy better than 1 \kms, as well as
orbits, have been computed for
all stars. They have been used to define 2 subsamples kinematically
representative of the thin disk and the thick
disk in order to highlight their respective properties. A transition
occurs at \FeH$=-0.3$. Stars more metal-rich than this value have a flat
distribution with \Zmax$<1$ kpc and
$\sigma_W<20$ \kms, and a narrow distribution of \alfe. There exist stars
in this metallicity regime which cannot belong to the thin disk because of
their
excentric orbits, neither to the thick disk because of their low scale height.
Several thin disk stars are identified down to \FeH$=-0.80$. Their Mg
enrichment is lower than thick disk stars with the same metallicity.
We confirm from a larger sample the results of Feltzing et al (\cite{felt03})
and Bensby et al (\cite{ben03})
showing a decrease of [$\alpha$/Fe] with \FeH in the
thick disk interpreted as the signature of the SNIa which have progressively
enriched the ISM with iron.
However our data suggest that the star formation in the thick disk stopped
when the enrichment was \FeH$=-0.30$, \MgFe$=+0.20$, \SiFe$=+0.17$. A
vertical gradient in [$\alpha$/Fe] may exist in the thick disk but should be
confirmed with a larger sample. Finally we have identified
2 new candidates of the HR1614 moving group.
}

\maketitle

\section{Introduction}
The behaviour of stellar elemental abundances with age and kinematics,
in various substructures of the Galaxy, puts important constraints to
the construction of models of chemical and dynamical evolution of our
galactic system.
In this paper we put particular attention to the transition between the thin disk and
the thick disk and  to the abundances of the $\alpha$-elements Mg and Si
and the iron-peak element Ni.
According to the current nucleosynthetic theory, $\alpha$-elements are being
produced
as a result of $\alpha$-capture reaction, taking place in the core of
massive stars during their explosion as SN II (Burbidge et
al \cite{bbfh57}).
Fe is produced by both massive SN II and less
massive SN Ia. It is expected that 2/3 of the solar abundance
of iron is produced by
explosions of white dwarfs in double systems (SN Ia), 1/3 by SN II
(Timmes et al 1995).
If the percentage of massive stars in the earlier
Galaxy was higher than today, one can predict that the $\alpha$/Fe
ratio is going to change over time. A well established observational fact
shows that in old metal-poor stars of the Galaxy, \alfe,
 in particular \MgFe, is overabundant relative to Sun's value
(Wallerstein 1961; Gratton \& Sneden 1987; Magain 1989; Nissen et al 1994;
Fuhrmann et al 1995 etc). If the $\alpha$-element overabundance is
a typical chemical pattern in halo stars in comparison with disk stars,
there is a question whether there is a distinction in
the Mg behaviour in other subsystems of the Galaxy, in particular in
the thin and thick disks. Such a difference would have important consequences
on the choice
of the most probable scenario of formation of the thick disk (collapse,
accretion etc.) and its timescale. Moreover \alfe could serve as an
additional criterion
for the deconvolution of thick and thin disk populations. It is especially
important for the stars with
metallicity in the interval of overlapping \FeH values
($-0.7<$\FeH$<-0.3$).

Several groups  have  recently attacked this problem by attempting
to make consistent analyses of abundances of Mg and other
elements in large ($>$ 50) samples of stars of the thin and thick disks, using
kinematics to distinguish the stellar populations. However, their results are
ambiguous.
Fuhrmann (1998) found the thick disk to be significantly older than the thin
disk and its data exhibit an
appreciable constant overabundance of MgI in the thick disk relative
to the thin disk. He interprets the observed discontinuity in the chemical and
kinematical
behaviour of the thin disk and thick disk populations as a hiatus in star
formation
before the earliest stages of the thin disk.
Mashonkina's group (Mashonkina \& Gehren 2000, Mashonkina \& Gehren 2001,
Mashonkina et al 2003) is in agreement with Fuhrmann concerning the
discontinuity between the thin
disk and the thick disk. They observe a step-like decrease of Eu/Ba ratio at
the thick disk
to thin disk transition and
estimate that the thick disk formed from well mixed gas during a short
timescale  of $\sim$ 0.5 Gyr.
Gratton et al (2000) argue in favour of a constant and significant
overabundance of Mg with
respect to Fe in the thick disk, a fast formation of the thick disk,
a sudden decrease in star formation during the transition between the thick
and thin disk phases and
a formation scenario mixing dissipational collapse and accretion
on similar time scales.
Contrary to these studies, Chen et al (2000) do not observe in their data any
discontinuity
between the thin disk and the thick disk. Their thick disk stars exhibit a
slightly higher  $\alpha$/Fe ratio than the thin disk,
but not exceeding +0.4.
Very recently Bensby et al (2003) and Feltzing et al (\cite{felt03})
published a convincing study showing that the thick disk extents
at solar metallicities with an abundance trend clearly separated from the thin
disk despite a
large overlap in metallicity.  A knee and decrease in \MgFe vs \FeH is
interpreted as the signature
of contribution from SNIa to the enrichment of the interstellar gas out of
which the thick disk stars formed.

However these studies cannot be easily compared because they do not rely on the
 same definition
of the thick disk. The thin disk and the thick disk are known to greatly
overlap in kinematics
and metallicity, and the selection of a sample representative of each
population  is not obvious.
The use of a single criterion (\Vrot, \Zmax or eccentricity
of the orbit, \FeH , ...) is usually not
sufficient for this task. In the last decade, studies of the thick disk
kinematics have been numerous
and its velocity distribution is now well constrained.
This allows to make a rigourous classification of the stellar populations by
computing the probability of each star of a sample to belong to the thin disk
and the thick disk
according to its velocity (U,V,W), having first evaluated the selection biases
which affect
local samples. It is the way we have adopted in this paper to investigate the
behaviour of
Mg and Si (as alpha-element) and Ni (as iron-peak element) in the thin and
thick disks.

In the past years, Mg abundances in stars have been determined in a wide range
of metallicities
and temperatures, either in LTE approximation or through
detailed NLTE calculations.
In the visible range, there are about 10 MgI lines,
but most of them are strong lines with equivalent width (EW) greater
than 200 m\AA, and for such lines NLTE effects can be significant.
As shown by several authors (Zhao \& Gehren 2000;
Idiart \& Th\'evenin 2000; Shymanskaya et al 2000), departures from LTE
are observed in MgI lines in stars of various spectral types.
We have therefore used the NLTE approach to determine Mg abundances more
exactly.
The lines that we have used for the other elements have small or moderate EW
and considering our temperature and metallicity interval, we have negleted NLTE corrections.
As a matter of fact, in the Sun the abundance deviations due to NLTE effects are
generally small and do not exceed 0.1 dex, as established in former
calculations: 0.07 dex according to Shchukina \& Bueno (2001), $<$0.10 dex
according to Gehren et al (2001a, 2001b) for FeI lines, and -0.01 dex
for SiI lines (Wedemeyer 2001). Gratton et al (1999) and
Th\'evenin \& Idiart (1999) have investigated NLTE effects
 for iron lines in dwarfs and giants of different
metallicity. In both papers, the dominating NLTE effect for Fe is the overionization by
ultraviolet radiation (UV photons).
Gratton et al (1999) found that NLTE corrections may be neglected
in most cases, including the stars on the main-sequence and red
giant branch. Th\'evenin \& Idiart (1999) derived metallicity corrections
of about 0.3 dex for stars with \FeH$<$--3.0, however for stars with
\FeH$>$--1 this value does not exceed 0.1 dex.
Abundances of Fe, Si and Ni were therefore determined in LTE approximation.

This paper presents the determination and analysis of kinematical parameters
and abundances of Mg, Si and Ni for 174 dwarf stars
in the domain of $-1.0<$\FeH$<+0.3$. This study was carried out using
a homogeneous spectral material, uniform methods of treatment and
NLTE calculations for 9 MgI lines. Two subsamples have been defined
on the basis of kinematics to be representative of the thin disk and the
thick disk and are used to highlight the chemical behaviour of the two
stellar populations.

\section{Observations and reduction}

All the spectra used in this paper are extracted from the most recent
version of the library
of stellar spectra collected  with the ELODIE echelle spectrograph
at the Observatoire de Haute-Provence by Soubiran et al (\cite{soub98})
and Prugniel \& Soubiran (\cite{pru01}). The performances of the instrument
mounted on the 193cm telescope, are described in Baranne et al (\cite{baran96}).
ELODIE is a very stable instrument, build to monitor radial
velocity oscillations of stars with exoplanets, at a resolving power
of 42\,000 in the  wavelength range $\lambda$ $\lambda$ 3850--6800 \AA \AA.
Spectrum extraction,
wavelength
calibration  and radial velocity measurement have been performed
at the telescope with the on-line data reduction software while
straightening of the orders, removing of cosmic ray hits, bad pixels and
telluric lines
were performed as described in Katz et al (1998).

All the spectra of the library have been parametrized in terms of (\Teff,
\logg, \FeH), either from the literature or internally with the TGMET
code (Katz et al 1998).  This allowed us to select a set of 174 FGK dwarfs
and subgiants (3.7$<$logg$<$4.6) for this study, with metallicities in the
range
$-1<$\FeH$<+0.3$. Several solar spectra taken on the Moon and asteroids
were also considered as references.
The selected spectra have a signal to noise ratio (S/N) at 5500 \AA \,
ranging from 100 to 300.

\begin{figure}[hbtp]
\begin{center}
\includegraphics[width=5cm]{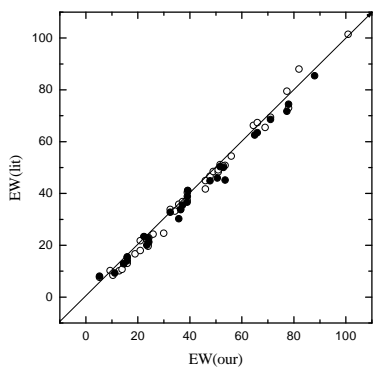}
\caption{Comparison of solar EW measured in this study
        with those from the literature (Edvardsson et al
        \cite{edv93} -- filled circles, Reddy et al \cite{red02} --
        open circles)}
\label{f:EWcomp}
\end{center}
\end{figure}

The continuum level drawing and equivalent width
measurements were carried out by us using DECH20 code (Galazutdinov, 1992).
Equivalent widths of lines were measured by Gaussian function fitting.
Their accuracy was estimated by comparing our
measurements on solar spectra to those obtained by other authors. The mean
difference with Edvardsson  et al (\cite{edv93}) is :
$\rm{EW}_{\rm Edv}-\rm{EW}_{\rm our}=-2.3 \,
(\sigma= 2.3$) m\AA \,  for 27 lines of FeI, FeII, SiI and NiI in common;
and with Reddy et al (\cite{red02}) :
$\rm{EW}_{\rm Red}-\rm{EW}_{\rm our}=-1.2 \, (\sigma= 2.1)$ m\AA \,
for 47 lines in common. The comparison is shown in Fig. \ref{f:EWcomp}.

\section{Atmospheric parameters}

In order to perform reliable abundance determinations, it is crucial to
derive accurate stellar parameters, especially effective temperatures \Teff.
V-K or IR flux methods are often advocated as the best (Asplund \cite{asp03})
but require homogeneous infrared observations which are not available for all stars.
Recently our group has improved the line-depth ratio
technique pioneered by Gray (1994), leading to high precision \Teff for
most of our program stars (Kovtyukh et al 2003).
This method, relying on ratios of the measured central depths of lines having
very different functional dependences on \Teff, is independent of
interstellar reddening and takes into account the individual
characteristics of the star's atmosphere.
Briefly, a set of 105 relations was obtained, the mean random error of a
single calibration being 60--70 K
(40--45 K in the most and 90--95 K in the least accurate
cases). The use of $\sim$70--100 calibrations per spectrum reduced the uncertainty to
5--7 K. These 105 relations were obtained from
92 lines, 45 with low ($\chi<$2.77 eV) and 47 with high ($\chi>$4.08 eV)
excitation potentials, calibrated from 45 reference stars in common
with Alonso et al (\cite{AAMR96}), Blackwell \& Lynas-Gray (\cite{BLG98}) and
di Benedetto (\cite{dB98}). The zero-point of
the temperature scale was directly adjusted to the Sun, based on 11 solar
reflection spectra taken with ELODIE, leading to the uncertainty
in the zero-point of about 1K. Judging by the small
scatter in our final calibrations and \Teff, the selected
combinations are only weakly sensitive to effects like rotation,
micro- and macroturbulence, non-LTE and other. The application range of the
line-depth method is $-0.5<$\FeH$<+0.5$.

\begin{figure}[hbtp]
\begin{center}
\includegraphics[width=6cm]{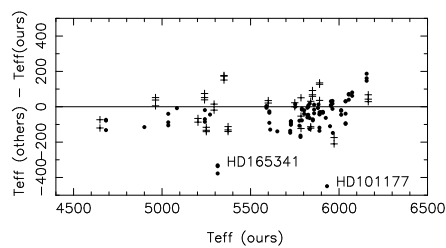}
\caption{Comparison of our adopted temperatures with those estimated
by Alonso et al (\cite{AAMR96}), Blackwell \& Lynas-Gray (\cite{BLG98})
and di Benedetto (\cite{dB98}).
Crosses indicate stars with \FeH$<$-0.5. The two outliers HD101177 and
HD165341 are discussed in the text.
HD165341 was measured by Alonso, Blackwell and Di Benedetto, that is why
there are 3 points for the same star.
\label{f:comp_teff}}
\end{center}
\end{figure}

For the most metal-poor stars, \Teff was determined earlier (Mishenina
\& Kovthyukh \cite{mk01}).
The H$_\alpha$ line-wing fitting was used for stars studed in this work.
 None of the stars from Mishenina \& Kovthyukh (2001) had their
temperature
measured by line depth ratio because they are too metal poor.
In order to testify that the temperature scales adopted in Mishenina
\& Kovthyukh (\cite{mk01}) and Kovtyukh et al (2003) are consistent, we show in
Fig. \ref{f:comp_teff} our adopted temperatures versus those estimated by Alonso et al
(\cite{AAMR96}), Blackwell \& Lynas-Gray (\cite{BLG98}) and di Benedetto (\cite{dB98}).
Our temperature scale is slightly hotter than their by $\sim 20-30$K, as mentioned in Kovtyukh
et al (2003), but the dispersion is satisfactory ($\sim 80$K, Tab. \ref{t:compare_teff}).
We have checked the 2 outliers HD101177
and HD165341. We are certain that Alonso et al's
temperature for HD101177 (5483K) is in error because recently Heiter \& Luck (\cite{heit03})
determined a temperature of 6000K similar to ours (5932K), a value confirmed by the H$_\alpha$
profile. HD165341 is a variable, active, rotating spectroscopic binary rending its photometric
measurements suspicious.
 On the basis of IR photometry,
Alonso et al
(\cite{AAMR96}), Blackwell \& Lynas-Gray (\cite{BLG98}) and di Benedetto
(\cite{dB98})
determined respectively 4978K, 4983K, 4937K.
Our determination of 5314K agrees well with that of Zboril \& Byrne
(\cite{zbo98}) who find \Teff=5260K.

\begin{table}
\caption[]{Mean difference and standard deviation between our temperatures and those estimated by Alonso et al
(\cite{AAMR96}), Blackwell \& Lynas-Gray (\cite{BLG98}) and di Benedetto (\cite{dB98})
for $\sim$ 40 stars in common.}
\label{t:compare_teff}
\begin{tabular}{lccc}
\hline
\hline
Reference & N & $\Delta T_{\rm eff} (K)$&$\sigma (K)$ \\
\hline
Alonso et al (\cite{AAMR96}) & 43 & 31 & 84 \\
Blackwell \& Lynas-Gray (\cite{BLG98}) & 41 & 29 & 84 \\
di Benedetto (\cite{dB98}) & 41 & 23 & 78 \\
\hline
\end{tabular}
\end{table}

\begin{figure}[hbtp]
\begin{center}
\includegraphics[width=7cm]{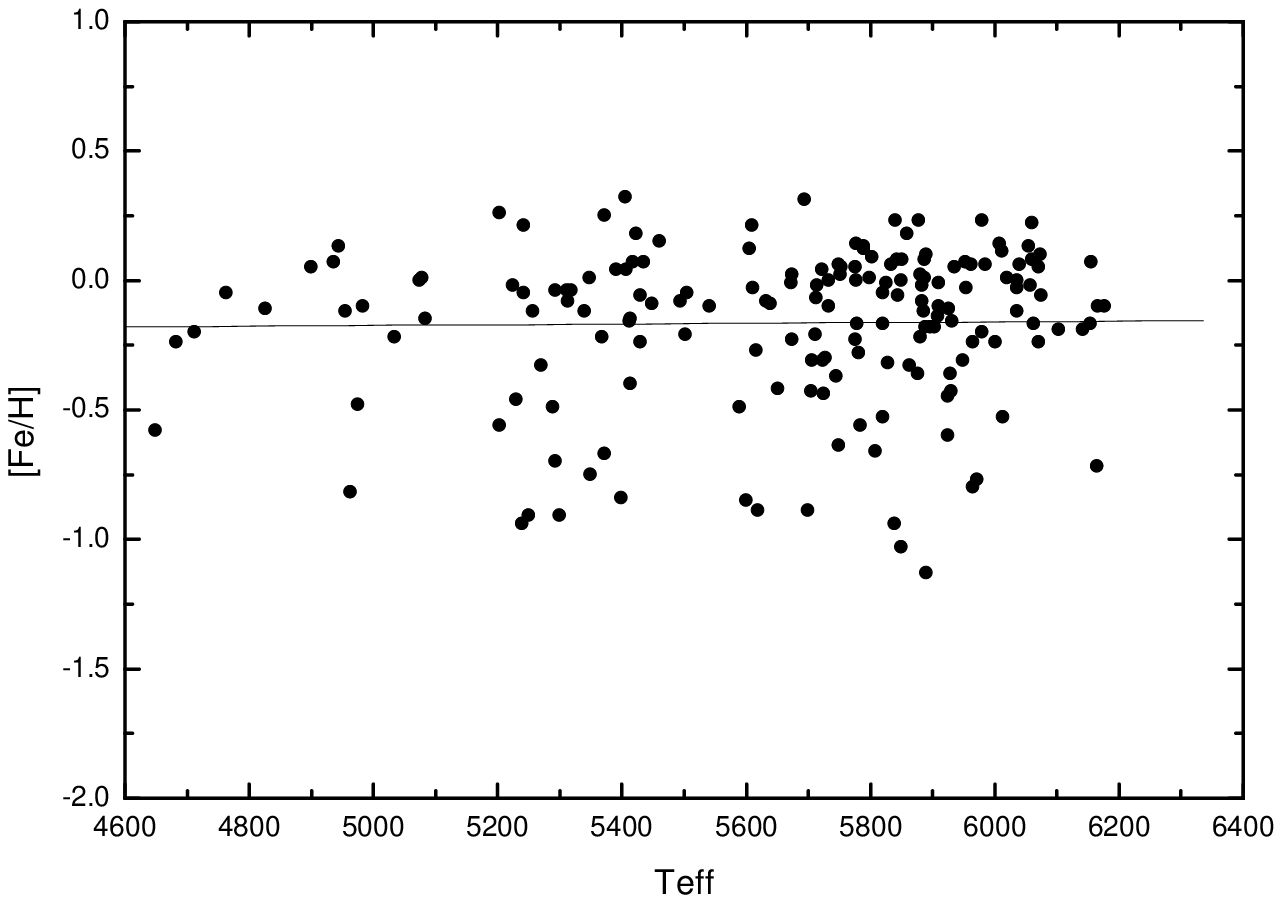}
\includegraphics[width=7cm]{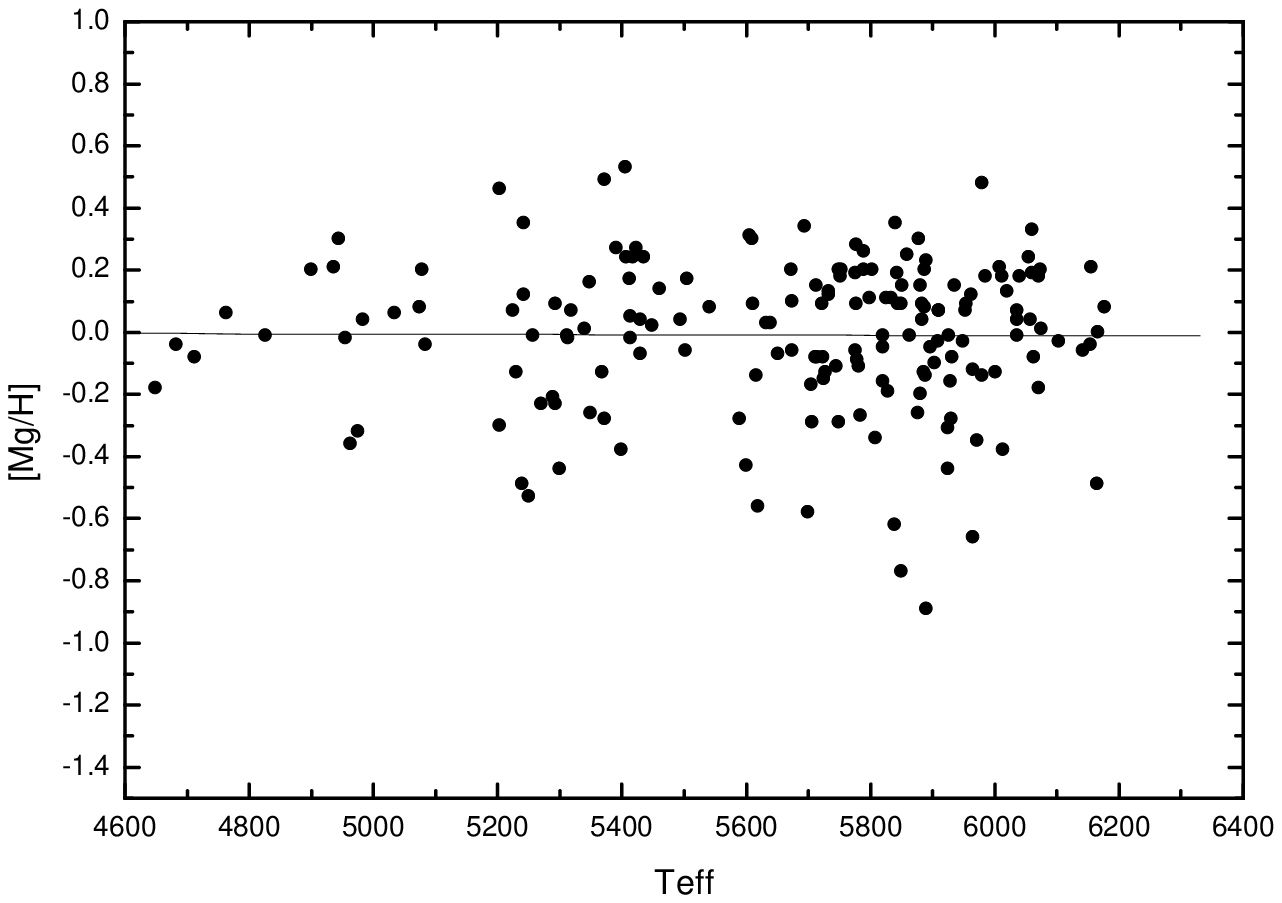}
\caption{Abundances of Fe and Mg vs \Teff}
\label{f:ab_tg}
\end{center}
\end{figure}

Finally, Fig. \ref{f:ab_tg} shows a general lack of correlation between
the Fe and Mg abundances
and \Teff that testifies to a correct choice of the effective temperatures.

The two most commonly used techniques to determine surface gravities are
the ionization balance of neutral and ionized species and the fundamental
relation which expresses the gravity as a function of mass, temperature and
bolometric absolute magnitude deduced from parallaxe. For the latter method,
stellar masses can be estimated
from evolutionary tracks, but metallicities and eventual $\alpha$-enhancement
have to be known first. We therefore choosed to derive \logg
from ionization balance, a method which might be affected by NLTE effects.
However, a detailed study of surface gravities derived by different
procedures was
preformed by Allende Prieto et al (\cite{all99}) who concluded that astrometric
and spectroscopic (iron ionization balance) gravities were in good agreement
in the metallicity range $-1.0<$\FeH$<+0.3$. We compared our adopted surface
gravities to those determined astrometrically
by Allende Prieto et al (\cite{all99})
and obtain a mean difference and standard deviation of
-0.01 and 0.15 respectively for 39 stars in common. This is consistent with
an accuracy of 0.1 dex on our estimated spectroscopic gravities.
As an additional criterion of the reliable choice of \logg we used
the wings-fitting for the MgIb lines. For all stars the difference does not
exceed 0.2 dex (such difference was detected only for a few program stars).

Microturbulent velocities $V_{\rm t}$ were determined by forcing the abundances
determined from individual FeI lines to be independent of equivalent width.
Starting with $V_{\rm t}$=1.1 \kms, we varied it until abundances
computed for FeI lines (20 m\AA $<$ EW $<$ 150 m\AA) and plotted as
a function of EW-s showed a zero slope. The precision of the
microturbulent velocity determination is 0.2 \kms.

The abundance of iron relative to solar one
\FeH=log(Fe/H)$_*$ -- log(Fe/H)$_\odot$
was used as the metallicity parameter of a star and was obtained
from FeI lines.

The adopted parameters of the target stars are given in Table \ref{t:all_param}.
To check the adopted model atmosphere parameters, we compared our values
with those derived by
Edvardsson et al (\cite{edv93}),
Gratton et al  (\cite{grat96})
Fuhrmann (\cite{fuhr98}),
Chen et al (\cite{chen00}) and Fulbright (\cite{ful00}).
The mean differences (other -- us) and standard deviations are given in Table
\ref{t:param_comp}.
On the whole, the agreement is rougly good.

\begin{table}
\caption[]{Number of stars in common, mean difference and standard deviation
of parameter comparison with 5 other studies.}
\label{t:param_comp}
\begin{tabular}{crcrcrcc}
\hline
\hline
N & $\Delta T_{\rm eff}$&$\sigma$&
 $\Delta\log\,g$&$\sigma$&$\Delta$\FeH&$\sigma$&Ref$^*$\\
\hline
37&  --2 &78&  0.03&  0.17  &--0.07&  0.08& 1 \\
39&  --28&82&  0.04&  0.22  &--0.04&  0.10& 2 \\
18&   10 &75&  0.04&  0.15  &--0.01&  0.09& 3 \\
17& --99 &56&--0.06&  0.13  &--0.08&  0.06& 4 \\
12& --15 &74&  0.05&  0.16  &  0.04&  0.09& 5 \\
\hline
\end{tabular}
\begin{itemize}
\item[*]
1 -- Edvardsson et al (1993);
2 -- Gratton et al (1996);
3 -- Fuhrmann (1998);
4 -- Chen et al (2000);
5 -- Fulbright (2000);
\end{itemize}
\end{table}

\section{Abundance determination}
\subsection{Fe, Si, Ni abundances}
Fe, Si, Ni abundances were determined from an LTE analysis of equivalent widths
using the WIDTH9 code and the atmosphere models by Kurucz (\cite{kurucz93}).
Appropriate models for each star were derived by means of standard interpolation
through \Teff  and \logg. The model metallicities were taken with
an accuracy  of $\pm$0.25 dex around the TGMET first guess.
Abundances of the investigated elements  Fe, Si, Ni were carried out
from a differential analysis relative to the Sun.
Solar abundances of Fe, Si and Ni were calculated with our solar EWs
and the oscillator strengths log\,gf from
Kovtyukh \& Andrievsky (1999), who derived log\,gf for 565 lines of the 27
chemical elements using solar spectra.
We determined : log A(Fe) = 7.55, log A(Si) = 7.55,
log A(Ni) = 6.25, where log(H) = 12 (according to Grevesse \& Sauval (1998)
log A(Fe) = 7.50, log A(Si) = 7.55, log A(Mg) = 7.58,
log A(Ni) = 6.25).

Depending on authors, the solar iron abundance
varies from 7.44 to 7.64 (Asplund et al 2000; Blackwell et al 1995;
Gehren et al 2000a; Grevesse \& Sauval 1998; Shchukina \& Bueno 2001, etc).
This disagreement is caused
by several factors: the systems of oscillators strengths,
the models of solar atmosphere (empirical, hydrodynamical, LTE or NLTE
assumptions), the solar spectrum  LTE or NLTE synthesis etc.
Using laboratory log gf, Blackwell et al (1995) have obtained
a solar iron abundance  of 7.64. From calculated and laboratory
log gf, Gehren (2001a) found an abundance 7.50 and 7.56 from
Fe II lines, and 7.47 and 7.56 from Fe I lines.
The detailed hydrodynamical models of
the Sun result in the solar value of iron abundance of 7.44 (Fe I lines) and
7.45 (Fe II lines) (Asplund et al \cite{asplundet00} 2000). In the same time of the meteoretic value
of the iron
abundance is 7.50 (Grevesse \& Sauval 1998). The same value was found for the
solar atmosphere by Shchukina \& Bueno (2001) using the NLTE approximation and
hydrodynamical solar model.
The solar silicon abundance obtained within the framework of the hydrodynamical
models (Asplund 2000) is sligtly lower (7.51) than the meteoric silicon
abundance (7.55) (Grevesse \& Sauval 1998).
To reduce the influence of the used systems of
oscillators strengths, atmosphere models and instrumental errors we used a
differential method of abundance  determination.

The list of Fe, Si and Ni lines, atomic parameters and solar EW are given in
Table \ref{t:lines}.
Relative abundances of Fe, Si and Ni are given in Table \ref{t:all_param}.

\begin{table}
\caption[]{List of Fe, Si and Ni lines, atomic parameters and solar EW. Only
available at the CDS}
\label{t:lines}
\end{table}

\subsection{NLTE calculations for Mg}

The determination of Mg abundance was carried out
through detailed NLTE calculations
using equivalent widths of 4 lines
($\lambda$ $\lambda$ 4730, 5711, 6318, 6319 \AA) and
profiles of 5 lines ($\lambda$ $\lambda$ 4571, 4703, 5172, 5183, 5528 \AA).

NLTE abundances of Mg were determined with the help of a modified version of
the MULTI code of Carlsson (\cite{Carl}) described in Korotin et al.
(\cite{kal99} \cite{kak99}).
We have included in the code opacity sources
from the ATLAS9 program (Kurucz \cite{Kur2}).  This enables a much more
accurate calculation of the continuum opacity and intensity distribution
in the UV region which is extremely important in the correct determination
of the radiative rates of $b-f$ transitions.
 For $b-b$ transitions  include only continuum opacity and
for $b-f$ transitions include continuum both opacities and line opacities
from ATLAS9.
A simultaneous solution of the radiative transfer and statistical
equilibrium equations has been performed in the approximation of complete
frequency redistribution for all lines.

We employed the model of magnesium atom consisting of 97 levels: 84 levels of
\ion{Mg}{i}, 12 levels of \ion{Mg}{ii} and a ground state of \ion{Mg}{iii}.
They were selected from works of Martin \& Zalubas (\cite{MarZal}) and
Biemont \& Brault (\cite{BiBr}).
A detailed structure of the multiplets was ignored and each $LS$ multiplet
was considered as a single term.
The fine structure was taken into account only for a level $3s3p^{3}P^{0}$,
since it is closely connected  to the most important transitions in magnesium
atom
( $\lambda$ $\lambda$ 2778 -- 2782\AA\AA~,
3829 -- 3838 \AA\AA~, 5167 \AA~, 5172 \AA~).

Within the described system of the magnesium atom levels, we considered the
radiative transitions between the first 59 levels of \ion{Mg}{i} and ground
level of \ion{Mg}{ii}. Transitions between the rest levels were not taken
into account and they were used only in the equations of particle number
conservation.

Only transitions with $\lambda<100000$ \AA~ were selected for the
analysis. All 424 $b-b$ transitions were included in the linearization
procedure.

\begin{figure}
\begin{center}
\caption{
Grotrian diagram for MgI. Displayed are the radiative transitions treated
explicity in the non-LTE calculations}
\label{f:mg_i}
\includegraphics[width=8cm]{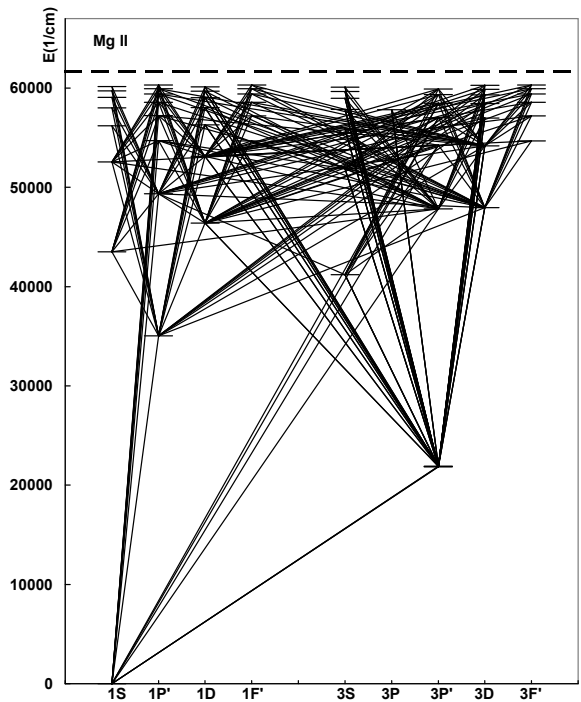}
\end{center}
\end{figure}

Grotrian diagram for the MgI are displayed in Fig. \ref{f:mg_i}.

Photoionization cross-sections were mainly taken from the Opacity Project
(Yan et al \cite{Yan}) keeping a detailed structure of their frequency
dependence, including resonances. For some important $b-f$ transitions,
the cross-section structure is extremely complicated making it difficult
to describe it using only simple approximations like $\sim \nu ^{-3}$.

Oscillator strengths were selected from the extensive compilative
catalogue by Hirata \& Horaguchi (\cite{Hir}). Some information was
obtained through the Opacity Project. As we ignored a multiple structure
of all the levels, the oscillator strengths for each averaged transition
were calculated as $f=\frac{\sum g_{i}f_{i}}{\sum g_{i}}$.

After the combined solution of radiative transfer and statistical
equilibrium equations, the averaged levels have been split with respect to
multiplet structure, then level populations were redistributed
proportionally to the statistical weights of the corresponding sublevels
and finally the lines of the interest were studied.

Electron impact ionization was described using Seaton's formula (Seaton
\cite{Seaton}). Collision rate between ground level \ion{Mg}{i} and ground
level of \ion{Mg}{ii} were approximated by use of the fits from Voronov
(\cite{Voronov}). For  electron impact excitation for all allowed $b-b$
transitions we used the van Regemorter (\cite{Regem}) formula. Collisional
rates for the forbidden transitions were calculated by using the semiempirical
formula (Allen \cite{Allen}), with a collisional force of = 1.

Inelastic collisions with hydrogen may play a significant role in the
atmospheres of cool stars. We took into account this effect with the help
of Drawin's formulas (\cite{Drawin}) offered Steenbock \& Holweger
(\cite{Steen}), with a correction factor of 1/3.
Nevertheless, as it was shown
in our test calculation, the influence of such collisions is negligible for the
stars considered in our paper. In particular, the variation of the correction
factor from 0 to 1 results in a change of the equivalent width less than 0.5\%.

For all the transitions
we also took into account such broadening parameters of lines as radiative
damping, Stark effect, van der Waals damping and microturbulent velocity. For
temperatures, considered by us, the influence of Stark effect is small.

The main influence on the profile was exerted by van der Waals broadening.
The Uns\"{o}ld's (\cite{Uns}) formula which is widely used to allow for
this effect, is known to yield somewhat understimated coefficients C$_{6}$. To
refine these coefficients, we conducted a special comparison of the observed
and calculated line profiles in solar spectrum using method Shimanskaya,
Mashonkina \& Sakhibullin (\cite{SMS}).
The Solar Flux Atlas of Kurucz et al. (\cite {Kur1}) and Kurucz's model
of the solar atmosphere (\cite{Kur2}) were used. To take into account the
chromospheric growth of temperature, this model was completed by a model of
the solar chromosphere from work of Maltby et al.(\cite{Maltb}). Nevertheless,
the influence of a chromosphere on the equivalent width of considered lines
was insignificant (less than 1\%).

The corrections {$\Delta$log C$_{6}$} to the classic Uns\"{o}ld (1955)
collisional damping constant derived from solar line wing fitting are
given in  Table \ref{t:mg}.
These values are comparable for common lines with results obtained in
the paper of Shimanskaya, Mashonkina \& Sakhibullin (\cite{SMS}).
Similar values were obtained by Barklem, Piskunov \& O'Mara
(\cite{BPO}). For example, their correction factors for the lines
$\lambda \lambda$ 5172, 5183 \AA \AA \,
are {$\Delta$log C$_{6}$}=0.85.
Oscillator strengths for observed lines were selected from the compilative
catalogue by Hirata \& Horaguchi (\cite{Hir}).

\begin{table}
\caption[]{Atomic data for the Mg lines}
\label{t:mg}
\begin{tabular}{rrrrc}
\hline
Lambda & $\chi$ (eV) & log\,gf & $\Delta$C$_{6}$ & $\Delta$C$_{6}$
(Barklem et al 2000)\\
\hline
4571.10&0.00&-5.79&1.72&\\
4702.99&4.35&-0.52&0.74&1.33\\
4730.03&4.35&-2.32&1.75&\\
5172.68&2.71&-0.38&0.74&0.85\\
5183.60&2.71&-0.16&0.72&0.85\\
5528.40&4.35&-0.49&0.44&0.95\\
5711.09&4.35&-1.72&0.60&\\
6318.72&5.11&-1.97&1.19&\\
6319.24&5.11&-2.18&1.19&\\
6319.49&5.11&-2.61&1.19&\\
\hline
\end{tabular}
\end{table}

The estimates obtained from the NLTE profile analysis of the MgI lines in the
solar
spectrum give the solar magnesium content (Mg/H)=7.57$\pm$0.02
which agrees well
with the determination of Grevesse \& Sauval (\cite {Grev98}) and
Shimanskaya, Mashonkina \& Sakhibullin (\cite{SMS}), who derived (Mg/H)=7.58.

\begin{figure}[hbtp]
\begin{center}
\includegraphics[width=8cm]{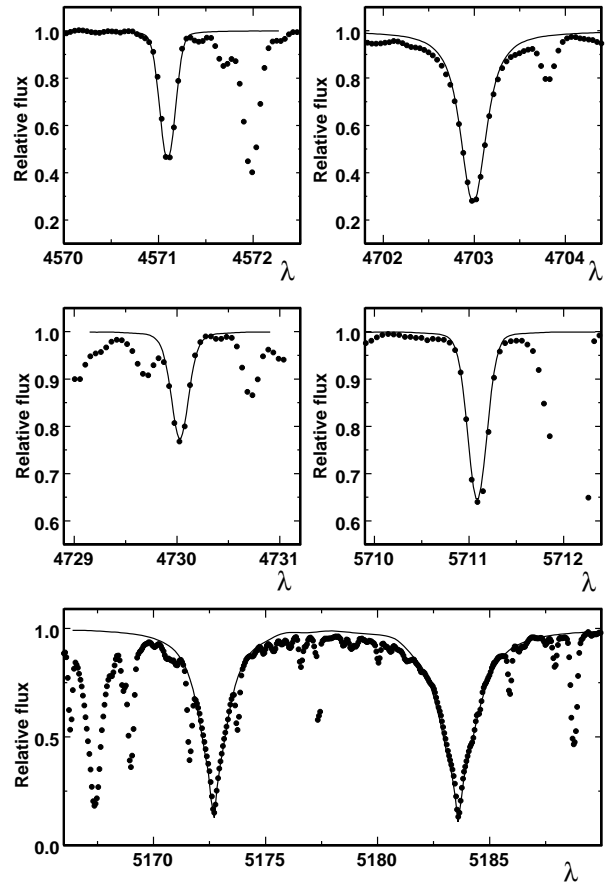}
\caption{Comparison of synthetic NLTE profiles (continuous line)
of the MgI lines to the observed spectrum of HD224930 (dots)
\label{f:NLTE_profile}}
\end{center}
\end{figure}

The comparison the NLTE profiles with those observed are presented in
Fig. \ref{f:NLTE_profile}. NLTE abundances of Mg are given in
Table \ref{t:all_param}.

The NLTE effects for the used lines appeared to be very small (less than
0.05 dex). The similar results were obtained in
Shimanskaya, Mashonkina \& Sakhibullin (\cite{SMS}).

\section{Error analysis}

Average standard deviations of abundances obtained from 72 - 265 lines
of Fe I (the number of used lines differs from star to star), 11 - 24 lines
of Si I and 12 - 17 lines of Ni I are 0.10, 0.08 and 0.08 respectively.
Final errors in abundances result mainly from errors in the choice
of parameters of the model atmospheres and in equivalent width
measurements (gaussian fitting, placement of the continuum) in the case of
Fe, Si, and Ni or fitting a synthetic spectrum in the case of Mg.
Table \ref{t:mod_errors} lists the errors obtained when changing the
atmospheric parameters by $\Delta$\Teff = -100 K;
$\Delta$\logg=+0.2; $\Delta$ V$_{t}$=+0.2 \kms;
$\Delta$\FeH=$\pm$0.25
and by assuming an uncertainty of $\pm$2 m\AA \, in the EW. These values were
adopted taking into account the intrinsic
accuracy of the atmospheric parameter determination, the processing
of the spectra
and the comparison of our parameter definition with those of other authors.
This test has been performed for 3 stars with different characteristics.

As seen in Table \ref{t:mod_errors}, the total uncertainty
reaches 0.12 dex for iron abundance determined from II species
and 0.10 dex in the case of I species.
The standard deviation obtained by comparing our \FeH determinations to
those obtained by other authors (Table \ref{t:param_comp}) shows that
we are consistent with them at a level lower than 0.10 dex.

\begin{table}
\caption[]{ Effects of uncertainties of model parameters and EW measurements on
the derived abundances}
\label{t:mod_errors}
\begin{tabular}{rrrrrrr}
\hline
\hline
\multicolumn{1}{c}{}&\multicolumn{2}{c}{HD004614}&
\multicolumn{3}{c}{5965/4.4/1.1/--0.24}&\multicolumn{1}{c}{}\\
\hline
El &  \Teff &    \logg &   Vt &    EW &   \FeH & Total \\
\hline
FeI &   0.07&   0.02&   0.04&  0.04&   0.00&   0.09 \\
FeII& --0.02& --0.08&   0.02&  0.04&   0.03&   0.10 \\
MgI &   0.05&   0.02&   0.02&  0.03&   0.00&   0.06 \\
SiI &   0.02&   0.01&   0.02&  0.03&   0.01&   0.05 \\
NiI &   0.07&   0.00&   0.03&  0.04&   0.00&   0.09 \\
\hline
\multicolumn{1}{c}{}&\multicolumn{2}{c}{HD022879}&
\multicolumn{3}{c}{5972/4.5/1.1/--0.77}&\multicolumn{1}{c}{}\\
\hline
FeI &   0.05&   0.02&   0.02&  0.08&   0.00&   0.10 \\
FeII& --0.01& --0.09&   0.01&  0.07&   0.02&   0.12 \\
MgI &   0.07&   0.03&   0.02&  0.04&   0.00&   0.09 \\
SiI &   0.02& --0.01&   0.00&  0.05&   0.00&   0.05 \\
NiI &   0.07&   0.01&   0.01&  0.06&   0.00&   0.09 \\
\hline
\multicolumn{1}{c}{}&\multicolumn{2}{c}{HD117176}&
\multicolumn{3}{c}{5611/4.0/1.0/-0.03}&\multicolumn{1}{c}{}\\
\hline
FeI &   0.07&   0.03&   0.04&  0.04&   0.01&   0.10 \\
FeII& --0.04& --0.10&   0.02&  0.04&   0.04&   0.12 \\
MgI &   0.06&   0.03&   0.03&  0.03&   0.00&   0.08 \\
SiI &   0.01& --0.01&   0.02&  0.03&   0.02&   0.05 \\
NiI &   0.06&   0.00&   0.03&  0.04&   0.01&   0.08 \\
\end{tabular}
\begin{list}{}{}\item[(1)] -- $\Delta$\Teff = -100 K;
\item[(2)] -- $\Delta$\logg= +0.2;
\item[(3)] -- $\Delta$V$_{t}$=+0.2 \kms;
\item[(4)] -- $\Delta$\FeH = --0.25 dex;
\item[(5)] -- $\Delta$EW= $\pm$ 2m\AA ~;
\item[(6)] -- Total error.
\end{list}
\end{table}

For Mg, Si and Ni, the total uncertainty due to parameter and EW errors
is 0.08 dex, 0.05 dex and 0.09 dex respectively.
We have also compared our Mg, Si and Ni abundances with those determined by
other authors.
For [Mg/H], we have carried out the comparison with Idiart \&
Th\'evenin (2000) and Gratton et al (2003) who have also used the NLTE approach.
Our [Si/Fe] and [Ni/Fe] determinations have been compared to those of
Edvardsson  et al (\cite{edv93}),
Reddy et al (\cite{red02}) and Chen et al (2000). The results of comparisons
are listed in Table \ref{t:abun_comp}. The mean differences are lower
than 0.05 and the standard deviations lower than 0.1 dex for all elements
proving the good consistency of our study with previous works relying on
similar methods.

\begin{table}
\caption[]{Abundance [Mg/H], [Si/Fe], [Ni/Fe] comparison with other authors}
\label{t:abun_comp}
\begin{tabular}{crcrcrcc}
\hline
\hline
N & $\Delta_{Mg}$ &$\sigma$&
 $\Delta_{Si}$&$\sigma$&$\Delta_{Ni}$&$\sigma$&Ref$^*$\\
\hline
18& 0.03 & 0.09 &&&&& 1\\
10& 0.04 & 0.09 &&&&& 2\\
37&  & & --0.05 & 0.05 & 0.01 & 0.05 & 3\\
17&  & & --0.04 & 0.05 & 0.02 & 0.03 & 4\\
5 & & & 0.01 & 0.08 & --0.05 & 0.09 & 5\\
\hline
\end{tabular}
\begin{itemize}
\item[*]
1 -- Idiard \& Th\'evenin (2000);
2 -- Gratton et al (2003);
3 -- Edvardsson et al (1993);
4 -- Chen et al (2000);
5 -- Reddy et al (2003)
\end{itemize}
\end{table}

\section{Stellar kinematics, metallicity, elemental abundances}

All the selected stars are bright and nearby enough to have parallaxes
and proper motions measured by Hipparcos (ESA \cite{ESA97}). These
astrometric quantities have
been combined with radial velocties measured on the ELODIE spectra
by cross-correlation (with an accuracy better than 100 $\mathrm{m\,s}^{-1}$) to compute
the 3 components (U,V,W) of the spatial velocities with respect
to the Sun. Combining the measurement errors on parallaxes, proper motions and
radial velocities, the resulting errors on velocities are of the order of 1 \kms.

Our first concern was to analyse the content of our sample in terms of stellar
populations. As we were interested in the characterization of abundance patterns
in the thin disk and the
thick disk, we have selected our sample to span the metallicity range $-1.0<$\FeH$<+0.3$
in order to define 2 subsamples representative of these two populations. A discrimination
of thin disk and thick disk stars is possible using the fact that
 the two disks are known to be distinct by their spatial distribution and local density, and by
their velocity, metallicity and age distributions. However we have performed this classification
with a pure kinematical appoach
because velocity distributions are the best constrained from observations reported in the literature.
If the thin disk velocity
distribution is well known thanks to Hipparcos data,  authors do not
yet agree on the properties of the thick disk.  A review of the current knowledge of the
thick disk is given  in Norris (\cite{N99}). There is
still a  controversy between the adherents  of a flat  and dense thick
disk, with typical scale height  of 800\,pc and local relative density
of 6-7\% (Reyl\'e \& Robin \cite{RR01}) to 15\%  (Soubiran et al \cite{soub03}),
 and the adherents of a thick disk with a higher
scale height,  typically 1300\,pc and a lower  local relative density,
of  the order of  2\% (Reid \& Majewski \cite{RM93}). Velocity dispersions are
generally found
to  span  typical  values  between 30  and  50  \kms,
 with an  asymmetric  drift  between  -20  and  -80
\kms. Recently, Soubiran et al (\cite{soub03}) determined the
kinematical parameters of the thick disk and the thick disk from an unbiased
sample of clump giants.
These values, listed in Tab. \ref{t:gauss}, are used in the present study.

\begin{table}
\caption[]{
\label{t:gauss}
Kinematical parameters adopted for
the thin disk and the thick disk from Soubiran et al (\cite{soub03}) }
\begin{tabular}{lcc}
\hline
\hline
 &  thin disk & thick disk   \\
\hline
$\rm V_{lag}$ (\kms) & -15 & -51 \\
$\sigma_{U} $ (\kms) & 39 & 63 \\
$\sigma_{V} $ (\kms) & 20 & 39 \\
$\sigma_{W} $ (\kms) & 20 & 39 \\
\hline
\end{tabular}
\end{table}

  The metallicity distributions of the thin and thick disks are also matter of debate.
Recent thick disk metallicity determinations quoted mean values from -0.36 (Bell \cite{bel96}) to
-0.70 (Robin et al \cite{rob96}, Gilmore et al \cite{gil95}), passing by -0.48 (Soubiran
et al \cite{soub03}). Morrison et al (\cite{MFF90})  brought to the fore low-metallicity stars
$(-1.6<$\FeH$<-1.0)$ with disk-like kinematics. Chiba \& Beers (\cite{CB00}) estimate
that 30\% of  the stars with $-1.6<$\FeH$<-1.0$ belong  to the thick
disk  population. Similarly, Bensby et al (\cite{ben03}) found in their sample stars with
super-solar metallicities and thick disk kinematics. It  remains  unclear whether  these
extreme populations  are separate  from  the thick  disk  or  their  metal-weak and metal-rich tails.
Moreover, due to the great overlap of the thin disk and thick disk distributions,
it is very difficult
to estimate where the transition occurs. Authors involved in the study of metallicity
distributions are often concerned with the chemical evolution of the Galaxy and generally do not
attempt to separate the thin and thick disks but rather consider the thick disk as an integral part
of the disk. It was the case for instance in the work of Edvardsson et al (\cite{edv93}) on
the age - metallicity relation. Haywood (\cite{H01}) performed a revision
of the solar neighbourhood metallicity distribution and concluded that it peaks at \FeH=0 and
that only 4\% of the nearby stars have \FeH$<-0.50$. These considerations made us think that
the selection of thin disk and thick disk stars on a pure kinematical criterion would be more
reliable.

\begin{figure}[hbtp]
\begin{center}
\includegraphics[width=8cm]{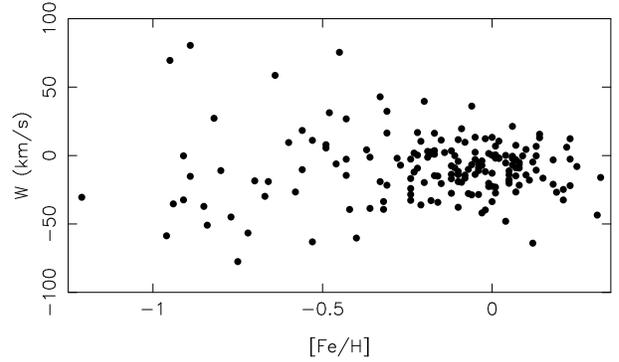}
\caption{
Distribution of the 174 stars in the plane metallicity - W velocity.
\label{f:FeH_W}}
\end{center}
\end{figure}

Fig. \ref{f:FeH_W} shows the distribution of the sample in the plane metallicity -
W velocity. It is clear from this plot that a transition occurs at \FeH$\sim$
-0.30 : above this value we measure a dispersion of the W velocities of 16 \kms
while below this value the dispersion is 38 \kms. These values are typical of
the thin disk and the thick disk respectively (see Tab \ref{t:gauss}). They are
related to the different scale heights of the two populations and testify that our sample
is indeed a mixture of thin and thick disk stars.

 In order to compute the
probability of each star to belong to the thin disk or the thick
disk on the basis of its spatial velocity, we need to know the kinematical
parameters
of the two disks and their proportion in our sample. The kinematical parameters are listed
in Tab. \ref{t:gauss}, but the proportions in our sample
are of course different of the real proportion in
the solar neibourhood because our selection in the range $-1<$\FeH$<+0.3$ is supposed to have
increased the number of thick disk stars comparatively to their relative
density in the solar neighbourhood (2\% to 15\%). Moreover the ELODIE
library was build from various observing programs which may have biased its content and
consequently our sample.

To estimate the proportion of thin and thick disk stars in our sample,
we have applied  an algorithm of deconvolution of multivariate
gaussian distributions, on (U,V,W) velocities. This algorithm, SEM
(Celeux \& Diebolt \cite{CD86}), solves
iteratively  the  maximum  likelihood  equations,  with  a
stochastic  step,  with no {\it a priori} information on the mixture of the
gaussian
distributions. The results are unambiguous, our sample is consistent with a
mixture of
2 gaussian populations with parameters very similar to those listed in Tab.
\ref{t:gauss},
but with relative densities of 75\% and 25\% respectively for the thin disk
and
the thick disk. Accordingly, we have computed  the
probability of each star, with a measured velocity (U,V,W), to belong to the
thin disk ($Pr_1$) and to the thick disk ($Pr_2$)  :

$$ Pr_1=p_1{F_1\over F},\,\, Pr_2=p_2{F_2\over F}$$
$$ F_i={1\over {\sqrt{2\pi}^3 \sigma_{U_i} \sigma_{V_i} \sigma_{W_i} }}
\exp{-0.5[{{U\over\sigma_{U_i}^2}+{(V-V_i)\over\sigma_{V_i}^2}+
{W\over\sigma_{W_i}^2}}]}$$
$$ F=p_1F_1+p_2F_2$$

\noindent where $p_1=0.75, p_2=0.25$ are the relative densities of the 2
populations in our sample and
$V_i,\sigma_{U_i},\sigma_{V_i},\sigma_{W_i},i=1,2$ are the kinematical
parameters of the thin
and thick disks listed in Table \ref{t:gauss}.

 From these probabilities we have selected two subsamples kinematically representative
 of each population
in order to highlight their respective properties.
Putting the limit at $Pr_i>80$\% ensures a minimal contamination of each subsample by
the other population. The thin disk subsample includes 109 stars whereas the thick disk subsample
include 30 stars. In the figures, unclassified stars
are represented by small dots,
thick disk stars by larger black dots, thin disk stars by open dots. The
only halo star of the sample,
HD194598 (\FeH=-1.21, V=-276.40 \kms) is not represented in the plots for
a better clarity.

\begin{figure}[hbtp]
\begin{center}
\includegraphics[width=8cm]{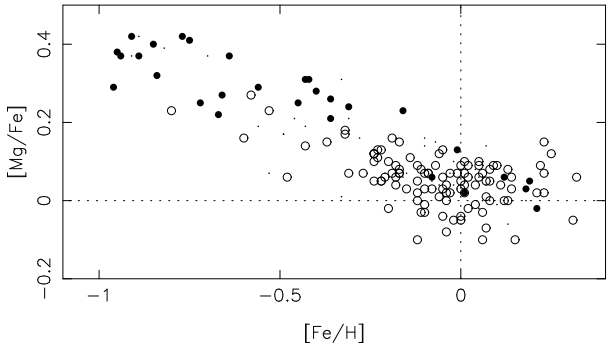}
\caption{
\label{f:mgfe_feh}
\MgFe vs \FeH for the whole sample. Black dots indicate
thick disk stars, open dots thin disk stars, small dots represent the
unclassified stars.}
\end{center}
\end{figure}

\begin{figure}[hbtp]
\begin{center}
\includegraphics[width=8cm]{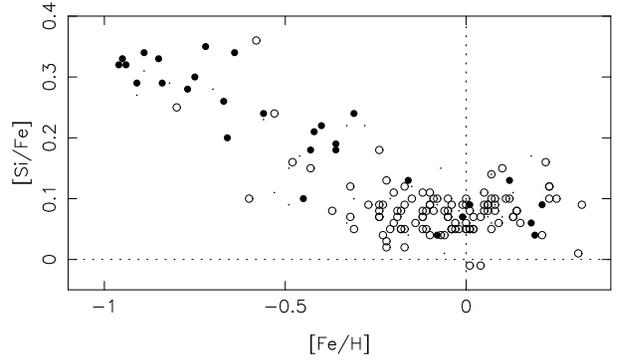}
\caption{
\label{f:sife_feh}
 Same as Fig. \ref{f:mgfe_feh} for Si.}
\end{center}
\end{figure}

\begin{figure}[hbtp]
\begin{center}
\includegraphics[width=8cm]{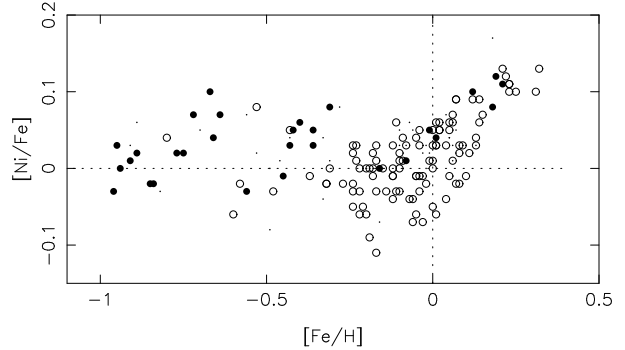}
\caption{
\label{f:nife_feh}
 Same as Fig. \ref{f:mgfe_feh} for Ni. }
\end{center}
\end{figure}

Fig. \ref{f:mgfe_feh}, \ref{f:sife_feh} and \ref{f:nife_feh} show the trend
of the abundances of Mg, Si and Ni as a function of \FeH. It can be seen
that :

\begin{itemize}
\item there are a few thin disk stars with \FeH$< -0.30$
\item metal poor stars (\FeH$<-0.60$) are all enriched in Mg and Si
(\MgFe$>+0.20$, \SiFe$>+0.15$)
\item on the contrary at solar metallicity, the enrichment of Mg and Si does
not exceed +0.20
\item at a given metallicity thick disk stars have higher \MgFe on average than
thin disk stars
\item the dispersion of \SiFe is remarkably small for \FeH$>-0.30$ but quite high
at lower metallicity
\item \MgFe and \SiFe decline with metallicity from about
+0.40,+0.35 to 0.0, +0.08 respectively
\item there are stars with thick disk kinematics at solar metallicities, their
abundance trends follow the thin disk
\item a rise and upturn is visible for \NiFe for \FeH$>0$
\end{itemize}

All these features are in very good agreement with those described in
 Bensby et al (\cite{ben03}).

\begin{figure}[hbtp]
\begin{center}
\includegraphics[width=8cm]{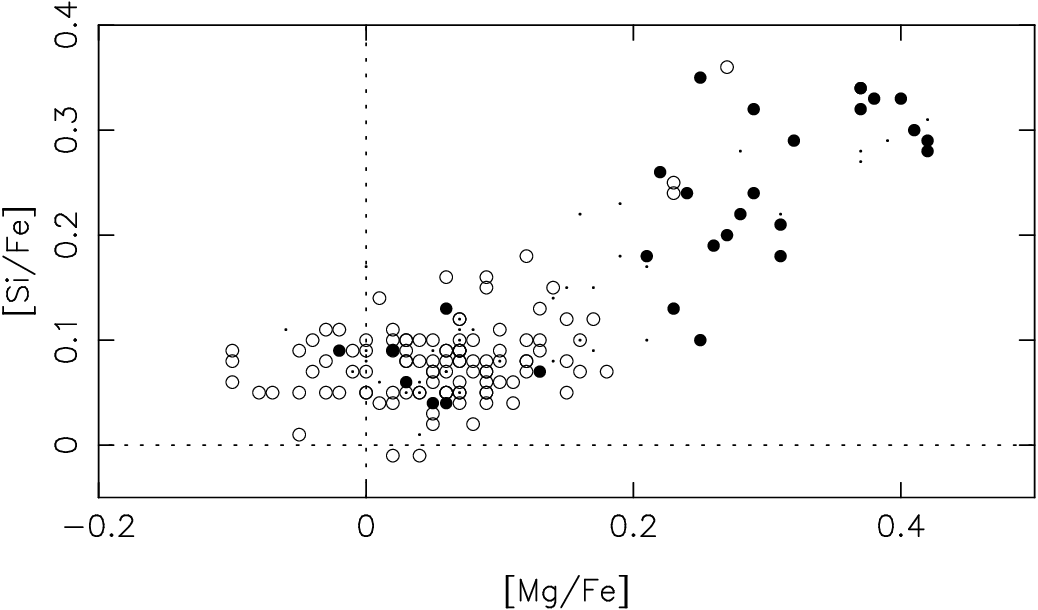}
\caption{
\label{f:sife_mgfe}
 Correlation between \MgFe and \SiFe}
\end{center}
\end{figure}

Fig. \ref{f:sife_mgfe} represents the correlation between Mg and Si relative
to iron. Mg and
Si are $\alpha$ elements which are supposed to be mainly produced in SNII.
It can
be seen in this plot that a transition occurs at \MgFe$\simeq+0.2$.
Stars with \MgFe$<+0.2$ have a mean abundance of \SiFe=+0.08 with a very
low dispersion of
$\pm 0.03$, lower than our error estimates.
\MgFe is more dispersed ($\pm 0.06$) around the mean of \MgFe=+0.05. For
stars with
\MgFe$>+0.2$ the distribution is consistent with a linear correlation :
\SiFe=0.7\MgFe+0.06 (rms=0.06).

\section{Discussion}

\subsection{\label{s:TDkin_Dmet}
Stars with thick disk kinematics and thin disk metallicity}
\begin{figure}[hbtp]
\begin{center}
\includegraphics[width=8cm]{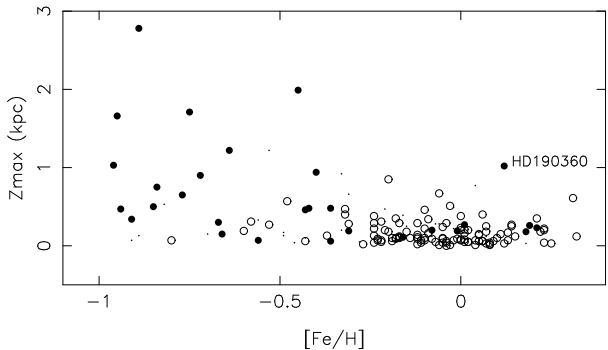}
\caption{
\label{f:feh_zmax}
The maximum distance to the plane of orbits, \Zmax, versus \FeH }
\end{center}
\end{figure}

A remarkable feature in the Fig. \ref{f:mgfe_feh} to \ref{f:sife_mgfe} is the
behavior of stars with
thick disk kinematics and thin disk metallicity. They follow exactly the
chemical trend of the thin
disk. This led us to wonder if these stars were real thick disk stars. According
to the W vs \FeH distribution shown in Fig. \ref{f:FeH_W},
metal-rich stars are expected to have a smaller scale height than metal poor
stars. To
have confirmation, we have computed the orbital parameters of our stars, by
integrating the
equations of motion in the
galactic model of Allen \& Santillan (\cite{allsan91}) over an age of 8
Gyr. The adopted velocity of the Sun with respect to the LSR is (9, 5, 6)
\kms, the solar galactocentric distance  ${\mathrm R}_{\odot}=8.5$ kpc
and circular velocity ${\mathrm V_{lsr}}=220$ \kms.
Table \ref{t:all_param} lists the spatial velocities  and the parameters of
the orbits : apogalactic and
perigalactic distances ($\mathrm{R_{min},R_{max}}$), the maximum distance from
the plane (\Zmax) and eccentricity ($e$). The distribution of \Zmax vs
\FeH in
Fig. \ref{f:feh_zmax} confirms the small scale height
of stars with thick disk kinematics and solar metallicity, ruling them out of
the thick disk. They
have been classified into the thick disk because of they have a large
eccentricity, but they have
a scale height which is typical of the thin disk. Thus
the origin of these stars cannot be in the thick disk neither in the local
thin disk
despite the same chemical behaviour. The origin of these stars has to be
found elsewhere.The only ambiguous star, HD190360 having \FeH=+0.12 and \Zmax=1 kpc,
appears as an outlier in the distribution. This star is discussed farther.

Among the 8 stars that verify
$\mathrm{Pr}_2>80$\% and \FeH$>-0.30$, HD030562, HD139323, HD139341 are K dwarfs with
very similar velocities and metallicities (Tab.
\ref{t:all_param}, $+0.18 \leq $\FeH$\leq +0.21$. HD030562 has been
identified to be part of the
HR 1614 moving group by Feltzing \& Holmberg (\cite{FH00}). It is very likely
that we have discovered
2 new candidates of this moving group whose age is estimated to be 2 Gyr. The
origin of this group is however unknown.

HD010145,  HD135204, HD152391 have $|U|>80$ \kms. An hypothesis is that they
have been thrown out from
the inner disk by the galactic bar. The over density of metal-rich stars with excentric orbits
confined near the plane have
have been interpreted as the signature of the galactic bar in the
solar neighbourhood by Raboud et al (\cite{rab98}) for instance.

Bensby et al (\cite{ben03}) have also found several stars with thick disk
kinematics and thin disk metallicity on the basis of a similar
probability calculation than ours, but using the ratio of $\mathrm{Pr_2}$ to $\mathrm{Pr_1}$.
If in our case these stars have a clearly a flatter distribution than the
thick disk, it was not mentioned in Bensby et al's study.  As it is important to
demonstrate if they belong or not to the thick disk we have calculated the orbits
of their stars and classified them with the same criterion as ours ($\mathrm{Pr_i}>80$\%).
The distribution of \Zmax versus \FeH is shown on Fig. 12.
Globally their distribution is similar to ours despite their sample is much smaller.
There is a sudden rise in \Zmax at \FeH$<-0.3$. They have also 3 outliers with
solar metallicities and \Zmax$\sim 1$ kpc, HD190360 being in common with us.
The five other stars with thick disk kinematics and \FeH$>-0.30$ have a
low scale height typical of the thin disk in agreement with our findings.
The question wether HD190360, HD003648 and HD145148 are really part of the
thick disk is an open question because they appear as outliers
in the distribution of kinematics vs metallicity. It is only with
complete samples with no kinematical bias that the relation of such
stars with the thick disk will
be clarified.

\begin{figure}[hbtp]
\begin{center}
\includegraphics[width=8cm]{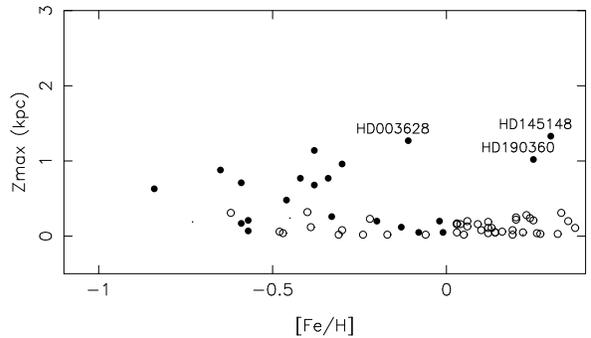}
\caption{
\label{f:feh_zmax_bensby}
The maximum distance to the plane of orbits, \Zmax, versus \FeH
for Bensby et al's sample.
The classification thin disk / thick disk is the same as ours
($\mathrm{Pr}_i>80$\%). }
\end{center}
\end{figure}

\subsection{The thin disk}
Several thin disk stars are found at low metallicity, down to \FeH$=-0.80$.
Several previous studies have also established the existence of such stars,
however the exact contribution of metal-poor stars to the thin disk
is difficult to evaluate from uncomplete samples.
Reddy et al (\cite{red02})
 have found in their samples a significant
number of stars with \FeH$<-0.40$
that they identify as belonging to the thin disk. They
have analysed 181 FG dwarfs, a sample size similar to ours. Their sample spans
the metallicity interval --0.70 to +0.15, with a peak at \FeH=--0.40. Only 3
stars were identified as belonging to the thick disk, based on the
criterion \MgFe$>+0.20$, all the other stars being identified as thin disk
stars. The peak at \FeH=--0.40 shows clearly that they have favoured in some way
the metal-poor tail of the thin disk. Unfortunatly their selection criteria
are not described.
On the contrary, the way Chen et al (\cite{chen00}) have favoured metal-poor
thin disk stars in their
sample is very clear : they have selected equal number of stars
in every metallicity interval of 0.1 dex in the range $-1.0<$\FeH$<+0.3$.
According to the low density of the thick disk in the solar neighbourhood,
thin disk stars are still dominating at moderatly low metallicity.
Haywood (\cite{H01}) has very carefully taken into account the selection biases
of his sample to study the metallicity distribution of the local disk but
his metallicities are photometric and no special attention was paid to a
kinematical distinction between the thin disk and the thick disk. Nevertheless
he found, contrary to Wyse \& Gilmore (\cite{WG95}), a low contribution of
the thin disk at \FeH$<-0.40$.

Our observations suggest that the distribution of \alfe in the thin
disk is very narrow, specially for Si, at \FeH$>-0.30$. On this
point we are in perfect agreement with Reddy et al (\cite{red02}) and Bensby
et al (\cite{ben03}).   For
the metal-rich part, we obtain as mean values and dispersions :
\MgFe=+0.05, $\sigma_{\mathrm
[Mg/Fe]}$=0.07, \SiFe=+0.07, $\sigma_{\mathrm [Si/Fe]}$=0.03. Such a narrow
chemical distribution implies that the stars formed from an homogeneous
gas. At lower metallicity, the Mg enhancement is higher. However at a given
metallicity the thin disk shows a lower Mg enhancement than the thick disk.
The behaviour of Si is quite surprising, showing high dispersion, whereas
Bensby et al (\cite{ben03}) find a similar behaviour than Mg. We notice also
that the Sun would be slightly Si-poor as compared to the thin disk. However
this offset stays within the error bars and might be related to the adopted zero
point of our abundance scales. Many previous studies (Edvardsson
et al \cite{edv93}, Chen et al \cite{chen00}, Reddy et al \cite{red02},
Bensby et al \cite{ben03}) exhibit also a positive offset in
\SiFe of the thin disk.

A rise of \NiFe
from $\sim$ --0.1 at \FeH=-0.20 to $\sim$ +0.1 at \FeH=+0.20 is visible in
Fig. \ref{f:nife_feh} as well as an upturn for the most metal rich star.
A similar feature is
also observed in Bensby et al (\cite{ben03}).

\subsection{The thick disk}
Having eliminated from the thick disk the 8 stars having high metallicity and
a flat distribution (Sect. \ref{s:TDkin_Dmet}),
we see from Figure \ref{f:mgfe_feh} that
\MgFe decreases from the halo value
(+0.40) at \FeH=-1.00 to +0.20 at \FeH=-0.30. Si has the same
behaviour, decreasing from +0.35
to +0.17, with one deviating star HD110897 having a lower enrichment in Si.
These findings are nicely consistent with the trends found
by Bensby et al (\cite{ben03}) that they interpreted as the signature of the
chemical enrichment by SNIa. Feltzing et al (\cite{felt03}) claimed the
existence of a "knee" in the \alfe abundance trends tracing the beginning of the
contribution of the SNIa. This "knee" may also exist in our data, especially in
\MgFe but the dispersion is too high to conclude. Definitively a larger sample
of thick disk stars is necessary to clarify this point. Contrary to
Bensby et al (\cite{ben03}) and Feltzing et al (\cite{felt03}), our data
suggest that the metallicity distribution of the thick disk does not
 continue at solar metallicity. The trend of declining ehancement, which was
also mentioned in Prochaska et al (\cite{pro00},
has to be compared to the results of Fuhrmann (\cite{fuhr98}) and Gratton et
al (\cite{grat00}) who found
in the thick disk a constant overabundance of $\alpha$-elements with respect
to iron, with a value similar
to that of the halo. In Fuhrmann (\cite{fuhr98}), only thin disk stars exhibit
a decrease of \MgFe with increasing \FeH. We believe that we have been able
to detect this trend,
as Feltzing et al
(\cite{felt03}) and Bensby et al (\cite{ben03}), because we have used a more
precise criterion to isolate a
sample a pure thick disk stars.
We agree with most of the previous works that a discontinuity exists between
the chemical distributions
of the thin and thick disks. Our data suggest that the star formation in the
thick disk stopped when
the gas had the composition \FeH=--0.3, \MgFe=+0.20, \SiFe=+0.17.

\begin{figure}[hbtp]
\begin{center}
\includegraphics[width=8cm]{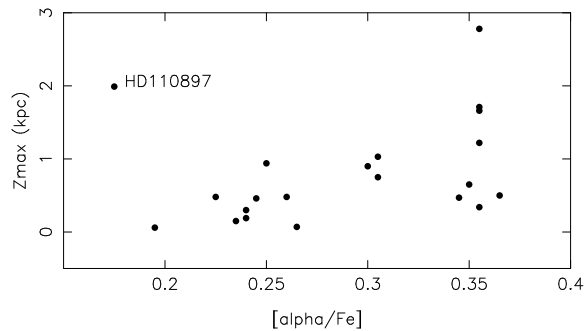}
\caption{
\label{f:alpha_zmax}
A possible vertical gradient in  [$\alpha$/Fe] in the thick disk ?}
\end{center}
\end{figure}

The distribution of \Zmax vs \FeH (Fig. \ref{f:feh_zmax}) does not exhibit a
clear vertical gradient in the thick disk metallicity. We have
searched for a vertical gradient in the $\alpha$-element abundances. Fig.
\ref{f:alpha_zmax} represents \Zmax vs \alfe=0.5(\MgFe+\SiFe) for
the thick disk (Pr2>0.80 and [Fe/H]<-0.3). A gradient
may exist, but should be confirmed with a larger sample. The existence of a
vertical gradient would imply
a significant timescale for the formation of the thick disk. This implication
is also true if the decline of $\alpha$ enhancement is interpreted by the
onset of SNIa.

\section{Summary and perspectives}
We have investigated the chemical behaviour of the thin disk and the thick disk
with a large sample of FGK dwarfs spanning the metallicity interval
[-1.00 ; +0.30]. We have used an homogeneous set of high resolution, high S/N
echelle spectra to derive abundances of Fe, Mg, Si and Ni with detailed
NLTE calculations
for Mg. All sources of errors have been analysed and we have compared our
abundances with several studies taken in the literature, showing a good
agreement. Orbits and space velocities were computed and used to estimate the
probability of each star to belong to the thin disk and the thick disk, taking
into account the selection bias that affects our local sample where the thick disk is
overrepresented as compared to the solar neighbourhood. Our main findings
concern the transition in kinematics and abundance trends which occurs at
\FeH$=-0.30$, the decline of $\alpha$
enhancement with increasing \FeH in the thick disk, and the narrow distribution
of Mg and Si in the thin disk. We have also observed
several stars which cannot belong to the thin disk
because of their large eccentricity, neither to the thick disk because of
their low scale height. This population, perhaps coming from the inner disk due
to a dynamical effect, has to be taken into account when studying the
metallicity distributions of the stellar populations in the solar neighbourhood.
Among these stars we have discovered two new candidates of the HR 1614 moving
group.

Larger samples are necessary to confirm our findings on the thin disk and the
thick disk. Unfortunatly thick disk stars are quite rare in the solar neighbourhood
and the construction of a significant sample of local thick disk stars implies either
a huge complete sample of disk stars, among which 2\% to 15\% of thick disk stars are expected,
 or the selection of such stars by kinematical or chemical criteria which
bias the conclusions which can be drawn. A lot of observing material is available.
In the public ELODIE
library, there are hundreds of high quality spectra of FGK stars that have not
been analysed yet. Most of these
stars have accurate parallaxes and proper motions from Hipparcos and their
radial velocities are also available. It
is thus possible to compute their probability to belong to a given kinematical
population. More difficult but necessary
is to evaluate the biases which render the metal-poor and high velocity stars
more represented in the library than in the solar
neighbourhood, reflecting the interest of observers in this kind of
stars. Another solution to avoid these biases is to observe the thick disk {\it in situ}
that is at distances and galactic latitudes where it begins to dominate the thin disk
in density. We are continuing such an observing program (Soubiran et al \cite{soub03}).

Detailed analysis of spectroscopic data is a tedious work
that takes much time to measure equivalent
widths, to fit profiles, to compute models etc... It is now absolutly necessary to develop automatic
methods to take advantage of the huge amount of high quality data which is
provided by the new generation of spectrographs.
We have undertaken such work to continue our investigation of the correlation
between elemental abundances and
kinematics among galactic disk stars from larger, more distant and complete
samples.

\begin{acknowledgements}
T.M. wants to thank the Observatoire de Bordeaux for kind hospitality.
Our special thanks to the referee Dr. M. Asplund for fruitful comments and suggestions.
This  research  has made use  of  the  SIMBAD and  VIZIER  databases,
operated at CDS (Strasbourg, France) and ESA products
(Hipparcos catalogue).
\end{acknowledgements}

\landscape
\begin{table}[!hb]
\caption[]{Atmospheric parameters, abundances, 3D velocities  with respect to
the Sun, orbital parameters, probability to belong to the thin  disk and the
thick disk for the 174 program stars}
\label{t:all_param}
\begin{tabular}{rrrrrrrrrrrrrrrr}
\hline
\hline
HD& \Teff & \logg &\FeH & \MgFe & \SiFe &   \NiFe & U & V& W & $\mathrm{R_{min}}$ & $\mathrm{R_{max}}$ &\Zmax &$\mathrm{ecc}$
& $\mathrm{Pr_1}$ & $\mathrm{Pr_2}$ \\
\hline
 HD000245   &5400&3.4&--0.84&+0.32&+0.29&--0.02& --44.40&--107.40& --50.90&  3.13&  8.70&  0.75&0.47&0.00&1.00\\
 HD001562   &5828&4.0&--0.32&+0.18&+0.07&--0.02&  17.90&  10.50& --33.60&  8.21& 10.33&  0.40&0.11&0.96&0.04\\
 HD001835   &5790&4.5&+0.13&+0.00&+0.07&+0.03& --35.70& --14.70&  --0.20&  7.40&  8.99&  0.07&0.10&0.98&0.02\\
 HD003765   &5079&4.3&+0.01&+0.02&+0.09&+0.04&  17.90& --82.50& --27.40&  4.06&  8.62&  0.27&0.36&0.04&0.96\\
 HD004307   &5889&4.0&--0.18&+0.06&+0.05&+0.00&  21.50& --25.10&   2.50&  6.82&  8.90&  0.10&0.13&0.97&0.03\\
 HD004614   &5965&4.4&--0.24&+0.10&+0.09&+0.02& --29.60& --10.90& --16.80&  7.72&  8.91&  0.13&0.07&0.97&0.03\\
 HD005294   &5779&4.1&--0.17&+0.08&+0.02&--0.03&  33.50&  --4.60& --14.70&  7.50&  9.85&  0.11&0.13&0.98&0.02\\
 HD006582   &5240&4.3&--0.94&+0.37&+0.32&+0.00& --42.40&--159.40& --35.30&  1.39&  8.61&  0.47&0.72&0.00&0.95\\
 HD008648   &5790&4.2&+0.12&+0.00&+0.10&+0.09&  --4.90& --15.80&  --4.70&  7.72&  8.53&  0.03&0.05&0.98&0.02\\
 HD009826   &6074&4.0&+0.10&+0.09&+0.15&--0.01&  28.70& --22.60& --14.20&  6.84&  9.13&  0.10&0.14&0.97&0.03\\
 HD010145   &5673&4.4&--0.01&+0.13&+0.07&+0.05&--112.20& --62.60& --20.30&  4.39& 10.44&  0.19&0.41&0.07&0.93\\
 HD010307   &5881&4.3&+0.02&+0.06&+0.05&+0.06& --37.90& --30.40&  --2.10&  6.55&  8.82&  0.05&0.15&0.95&0.05\\
 HD010476   &5242&4.3&--0.05&+0.03&+0.08&--0.01&  34.70& --24.90&   2.40&  6.62&  9.24&  0.10&0.17&0.96&0.04\\
 HD010780   &5407&4.3&+0.04&+0.04&--0.01&+0.03& --24.60& --16.40&  --5.40&  7.52&  8.70&  0.01&0.07&0.98&0.02\\
 HD011007   &5980&4.0&--0.20&+0.10&+0.06&+0.00&  25.40&  17.90&  39.60&  8.19& 11.37&  0.85&0.16&0.92&0.08\\
 HD013403   &5724&4.0&--0.31&+0.24&+0.24&+0.08& --22.20& --67.00& --21.70&  4.78&  8.57&  0.19&0.28&0.34&0.66\\
 HD013507   &5714&4.5&--0.02&--0.05&+0.05&--0.02&  --9.30&  --7.00& --11.80&  8.37&  8.52&  0.07&0.01&0.98&0.02\\
 HD013783   &5350&4.1&--0.75&+0.41&+0.30&+0.02&  48.60&   6.40& --77.50&  7.67& 11.70&  1.71&0.21&0.14&0.86\\
 HD013974   &5590&3.8&--0.49&+0.17&+0.09&--0.08& --40.40& --44.00&   7.90&  5.82&  8.77&  0.17&0.20&0.87&0.13\\
 HD014374   &5449&4.3&--0.09&+0.07&+0.08&+0.01& --18.20&  --8.30& --19.50&  8.13&  8.69&  0.16&0.03&0.98&0.02\\
 HD017674   &5909&4.0&--0.14&+0.11&+0.06&--0.03& --24.70& --18.80&   2.20&  7.40&  8.71&  0.10&0.08&0.98&0.02\\
 HD017925   &5225&4.3&--0.04&+0.04&+0.05&+0.03& --15.20& --21.90&  --9.00&  7.28&  8.53&  0.04&0.08&0.97&0.03\\
 HD019019   &6063&4.0&--0.17&+0.15&+0.05&--0.11& --40.10& --15.10&   3.60&  7.32&  9.12&  0.12&0.11&0.97&0.03\\
 HD019308   &5844&4.3&+0.08&+0.00&+0.09&+0.03& --54.70& --43.60& --21.70&  5.73&  9.08&  0.19&0.23&0.78&0.22\\
 HD019373   &5963&4.2&+0.06&--0.03&+0.08&+0.06& --75.50& --15.60&  21.30&  6.64& 10.34&  0.40&0.22&0.91&0.09\\
 HD022049   &5084&4.4&--0.15&+0.03&+0.10&+0.00&  --3.60&   7.00& --20.50&  8.48&  9.54&  0.18&0.06&0.98&0.02\\
 HD022484   &6037&4.1&--0.03&+0.02&+0.05&--0.01&   1.10& --15.30& --41.90&  7.79&  8.62&  0.51&0.05&0.91&0.09\\
 HD022556   &6155&4.2&--0.17&+0.07&+0.12&+0.03&   7.10&  10.20&   0.90&  8.41& 10.01&  0.10&0.09&0.98&0.02\\
\hline
\end{tabular}
\end{table}

\begin{table}[!hb]
\begin{tabular}{rrrrrrrrrrrrrrrr}
\hline
\hline
HD& \Teff & \logg &\FeH & \MgFe & \SiFe &   \NiFe & U & V& W & $\mathrm{R_{min}}$ & $\mathrm{R_{max}}$ &\Zmax &$\mathrm{ecc}$
& $\mathrm{Pr_1}$ & $\mathrm{Pr_2}$ \\
\hline
 HD022879   &5972&4.5&--0.77&+0.42&+0.28&+0.02&--109.60& --85.60& --44.90&  3.66& 10.05&  0.65&0.47&0.00&1.00\\
 HD023050   &5929&4.4&--0.36&+0.21&+0.18&+0.03& --68.20& --66.70&  --1.20&  4.58&  9.17&  0.06&0.33&0.26&0.74\\
 HD024053   &5723&4.4&+0.04&--0.01&+0.07&--0.04&  --4.80& --10.90&   0.10&  8.07&  8.56&  0.07&0.03&0.98&0.02\\
 HD024206   &5633&4.5&--0.08&+0.05&+0.09&+0.03& --14.40& --47.00& --16.50&  5.79&  8.53&  0.12&0.19&0.85&0.15\\
 HD028005   &5980&4.2&+0.23&+0.15&+0.12&+0.10& --46.70& --22.10& --22.00&  6.88&  9.18&  0.20&0.14&0.94&0.06\\
 HD028447   &5639&4.0&--0.09&+0.07&+0.10&+0.04& --28.80&  --6.10&  19.60&  8.01&  9.12&  0.34&0.06&0.97&0.03\\
 HD029150   &5733&4.3&+0.00&+0.09&+0.05&+0.00&   6.50&  --7.50&   0.20&  8.03&  8.89&  0.07&0.05&0.98&0.02\\
 HD029310   &5852&4.2&+0.08&+0.00&+0.09&--0.02& --45.40& --15.50&  --5.00&  7.22&  9.29&  0.02&0.13&0.97&0.03\\
 HD029645   &6009&4.0&+0.14&+0.03&+0.08&+0.06& --55.60& --20.40&  13.00&  6.81&  9.48&  0.25&0.16&0.95&0.05\\
 HD030495   &5820&4.4&--0.05&+0.06&+0.07&--0.01& --23.80&  --8.50&  --2.90&  7.98&  8.81&  0.04&0.05&0.98&0.02\\
 HD030562   &5859&4.0&+0.18&+0.03&+0.06&+0.08& --52.00& --72.80& --21.00&  4.41&  8.83&  0.18&0.33&0.13&0.87\\
 HD032147   &4945&4.4&+0.13&--0.06&+0.11&+0.03&   5.10& --54.70& --10.70&  5.36&  8.55&  0.05&0.23&0.76&0.24\\
 HD033632   &6072&4.3&--0.24&+0.12&+0.07&--0.05&   1.50&  --3.40& --24.10&  8.32&  8.93&  0.22&0.04&0.98&0.02\\
 HD034411   &5890&4.2&+0.10&--0.01&+0.07&+0.03& --75.30& --35.10&   4.40&  5.86&  9.72&  0.13&0.25&0.86&0.14\\
 HD038858   &5776&4.3&--0.23&+0.11&+0.04&+0.01& --17.20& --29.60& --12.10&  6.78&  8.54&  0.07&0.11&0.96&0.04\\
 HD039587   &5955&4.3&--0.03&+0.07&+0.06&--0.07&  12.20&   1.60&  --7.20&  8.15&  9.47&  0.01&0.07&0.98&0.02\\
 HD040616   &5881&4.0&--0.22&+0.05&+0.02&--0.03& --85.10&   5.30&  --9.20&  7.13& 11.81&  0.05&0.25&0.94&0.06\\
 HD041330   &5904&4.1&--0.18&+0.04&+0.08&--0.01&  10.20& --25.80& --32.90&  7.00&  8.71&  0.35&0.11&0.92&0.08\\
 HD041593   &5312&4.3&--0.04&--0.08&+0.05&--0.03&  10.70&   0.30& --11.00&  8.17&  9.36&  0.06&0.07&0.98&0.02\\
 HD043587   &5927&4.1&--0.11&+0.08&+0.08&+0.06& --18.30&  15.70&  --8.90&  8.48& 10.35&  0.04&0.10&0.98&0.02\\
 HD043856   &6143&4.1&--0.19&+0.07&+0.08&+0.00&  62.40&  --0.70&  --1.30&  7.04& 11.29&  0.06&0.23&0.97&0.03\\
 HD043947   &6001&4.3&--0.24&+0.12&+0.08&--0.03& --38.90& --11.20&  --2.60&  7.54&  9.20&  0.04&0.10&0.98&0.02\\
 HD045067   &6058&4.0&--0.02&+0.04&+0.05&     & --17.60& --65.90&  12.30&  4.84&  8.54&  0.22&0.28&0.45&0.55\\
 HD050281   &4712&3.9&--0.20&--0.02&+0.11&--0.06&   0.10&  12.80& --19.70&  8.47& 10.10&  0.18&0.09&0.98&0.02\\
 HD051419   &5746&4.1&--0.37&+0.15&+0.08&--0.01&  25.60&  14.20&   4.20&  8.10& 10.80&  0.13&0.14&0.98&0.02\\
 HD055575   &5949&4.3&--0.31&+0.21&+0.10&+0.01& --79.70&  --1.70&  32.30&  7.06& 11.26&  0.66&0.23&0.88&0.12\\
 HD058595   &5707&4.3&--0.31&+0.07&+0.05&+0.00& --10.80&  --7.70&  16.40&  8.36&  8.55&  0.28&0.01&0.98&0.02\\
 HD061606   &4956&4.4&--0.12&--0.10&+0.08&--0.04&  25.50&  --2.30&  --7.50&  7.73&  9.69&  0.02&0.11&0.98&0.02\\
\hline
\end{tabular}
\end{table}

\begin{table}[!hb]
\begin{tabular}{rrrrrrrrrrrrrrrr}
\hline
\hline
HD& \Teff & \logg &\FeH & \MgFe & \SiFe &   \NiFe & U & V& W & $\mathrm{R_{min}}$ & $\mathrm{R_{max}}$ &\Zmax &$\mathrm{ecc}$
& $\mathrm{Pr_1}$ & $\mathrm{Pr_2}$ \\
\hline
 HD062613   &5541&4.4&--0.10&+0.07&+0.04&+0.02&  --8.80&   8.70& --37.80&  8.51&  9.77&  0.46&0.07&0.95&0.05\\
 HD064606   &5250&4.2&--0.91&+0.37&+0.27&+0.03& --77.50& --60.90&  --0.20&  4.75&  9.39&  0.07&0.33&0.35&0.65\\
 HD064815   &5864&4.0&--0.33&+0.31&+0.22&+0.04&  72.90& --32.10&  42.90&  5.88& 10.55&  0.92&0.28&0.60&0.40\\
 HD065583   &5373&4.6&--0.67&+0.22&+0.26&+0.10& --13.20& --89.10& --29.80&  3.82&  8.52&  0.30&0.38&0.01&0.99\\
 HD065874   &5936&4.0&+0.05&+0.09&+0.08&+0.06&   9.00&  --4.10& --17.60&  8.10&  9.16&  0.14&0.06&0.98&0.02\\
 HD066573   &5821&4.6&--0.53&+0.23&+0.24&+0.08&  66.40&  25.90&  11.10&  7.53& 13.44&  0.27&0.28&0.94&0.06\\
 HD068017   &5651&4.2&--0.42&+0.31&+0.21&+0.05& --47.80& --60.60& --39.40&  5.03&  8.83&  0.48&0.27&0.22&0.78\\
 HD068638   &5430&4.4&--0.24&+0.05&+0.07&--0.02& --49.30& --18.90& --28.40&  6.99&  9.32&  0.29&0.14&0.93&0.07\\
 HD070923   &5986&4.2&+0.06&+0.07&+0.09&+0.07&  23.00& --37.30&  --0.50&  6.16&  8.85&  0.06&0.18&0.94&0.06\\
 HD071148   &5850&4.2&+0.00&+0.03&+0.05&+0.01&  20.10& --38.20& --22.50&  6.16&  8.78&  0.20&0.18&0.90&0.10\\
 HD072760   &5349&4.1&+0.01&+0.02&--0.01&--0.06& --35.00& --18.90&  --1.90&  7.22&  8.89&  0.05&0.10&0.97&0.03\\
 HD072905   &5884&4.4&--0.07&+0.09&+0.04&--0.04&  10.30&   0.40& --10.10&  8.18&  9.34&  0.05&0.07&0.98&0.02\\
 HD073344   &6060&4.1&+0.08&+0.08&+0.10&+0.03&  --4.90& --24.60&  --9.10&  7.13&  8.54&  0.04&0.09&0.97&0.03\\
 HD075732   &5373&4.3&+0.25&+0.12&+0.10&+0.10& --36.90& --18.10&  --8.00&  7.22&  8.95&  0.03&0.11&0.97&0.03\\
 HD076151   &5776&4.4&+0.05&+0.06&+0.09&+0.05& --40.30& --20.10& --11.20&  7.07&  9.00&  0.06&0.12&0.97&0.03\\
 HD076932   &5840&4.0&--0.95&+0.38&+0.33&+0.03& --48.10& --90.30&  69.50&  3.99&  8.80&  1.66&0.38&0.00&1.00\\
 HD081809   &5782&4.0&--0.28&+0.16&+0.22&+0.08& --42.80& --46.90&  --2.00&  5.66&  8.78&  0.05&0.22&0.84&0.16\\
 HD082106   &4827&4.1&--0.11&--0.03&+0.05&--0.03& --40.80& --13.70&  --0.10&  7.37&  9.16&  0.07&0.11&0.98&0.02\\
 HD088072   &5778&4.3&+0.00&--0.05&+0.09&+0.03& --21.20&   4.00& --33.70&  8.40&  9.47&  0.38&0.06&0.96&0.04\\
 HD089251   &5886&4.0&--0.12&+0.05&+0.07&+0.03&  19.70&  --3.80& --17.00&  7.85&  9.45&  0.14&0.09&0.98&0.02\\
 HD089269   &5674&4.4&--0.23&+0.13&+0.09&+0.03&  12.90& --28.00&   1.80&  6.77&  8.71&  0.09&0.12&0.97&0.03\\
 HD091347   &5931&4.4&--0.43&+0.14&+0.15&+0.05&  50.40&  27.90&  --2.70&  7.85& 12.87&  0.06&0.24&0.96&0.04\\
 HD095128   &5887&4.3&+0.01&+0.07&+0.05&+0.03& --24.00&  --2.50&   0.50&  8.20&  9.06&  0.08&0.05&0.98&0.02\\
 HD098630   &6060&4.0&+0.22&+0.09&+0.16&+0.12&  19.00& --25.50&   6.10&  6.83&  8.89&  0.18&0.13&0.97&0.03\\
 HD101177   &5932&4.1&--0.16&+0.06&+0.04&--0.07& --51.70& --27.20& --34.20&  6.58&  9.20&  0.39&0.17&0.86&0.14\\
 HD102870   &6055&4.0&+0.13&+0.08&+0.07&+0.04&  40.30&   3.40&   6.70&  7.56& 10.53&  0.17&0.16&0.98&0.02\\
 HD106516   &6165&4.4&--0.72&+0.25&+0.35&+0.07&  54.50& --75.60& --56.60&  4.29&  9.20&  0.90&0.36&0.01&0.99\\
\hline
\end{tabular}
\end{table}

\begin{table}[!hb]
\begin{tabular}{rrrrrrrrrrrrrrrr}
\hline
\hline
HD& \Teff & \logg &\FeH & \MgFe & \SiFe &   \NiFe & U & V& W & $\mathrm{R_{min}}$ & $\mathrm{R_{max}}$ &\Zmax &$\mathrm{ecc}$
& $\mathrm{Pr_1}$ & $\mathrm{Pr_2}$ \\
\hline
 HD107213   &6156&4.1&+0.07&+0.14&+0.14&+0.02& --25.70& --48.60& --14.90&  5.67&  8.58&  0.12&0.20&0.82&0.18\\
 HD107705   &6040&4.2&+0.06&--0.10&+0.09&+0.01& --16.10& --19.30&  --1.10&  7.44&  8.53&  0.06&0.07&0.98&0.02\\
 HD108076   &5700&4.3&--0.89&+0.42&+0.31&+0.06& --93.90& --41.70& --15.20&  5.36& 10.19&  0.13&0.31&0.62&0.38\\
 HD108954   &6037&4.4&--0.12&+0.09&+0.07&--0.01&  --0.20&   8.60& --27.60&  8.46&  9.75&  0.29&0.07&0.97&0.03\\
 HD109358   &5897&4.2&--0.18&+0.09&+0.05&+0.02& --30.80&  --3.50&   1.20&  7.99&  9.20&  0.09&0.07&0.98&0.02\\
 HD110833   &5075&4.3&+0.00&--0.04&+0.10&+0.05& --19.00& --21.40&  13.30&  7.30&  8.57&  0.23&0.08&0.97&0.03\\
 HD110897   &5925&4.2&--0.45&+0.25&+0.10&--0.01& --41.70&   6.90&  75.40&  8.14& 11.05&  1.99&0.15&0.20&0.80\\
 HD112758   &5203&4.2&--0.56&+0.19&+0.23&+0.07& --74.90& --35.20&  18.30&  5.87&  9.70&  0.33&0.25&0.82&0.18\\
 HD114710   &5954&4.3&+0.07&--0.07&+0.05&--0.02& --50.40&  11.50&   7.50&  7.91& 10.79&  0.18&0.15&0.97&0.03\\
 HD115383   &6012&4.3&+0.11&+0.04&+0.10&+0.02& --38.60&   1.50& --18.00&  7.95&  9.72&  0.15&0.10&0.97&0.03\\
 HD116443   &4976&3.9&--0.48&+0.06&+0.16&--0.03&   1.50&   5.20&  31.30&  8.41&  9.59&  0.57&0.07&0.96&0.04\\
 HD117043   &5610&4.5&+0.21&+0.02&+0.04&+0.13& --35.40& --26.80& --32.50&  6.82&  8.82&  0.35&0.13&0.90&0.10\\
 HD117176   &5611&4.0&--0.03&+0.07&+0.05&+0.02&  13.00& --51.90&  --3.90&  5.46&  8.61&  0.03&0.22&0.82&0.18\\
 HD117635   &5230&4.3&--0.46&+0.21&+0.17&+0.01&--137.80& --30.30&  --6.10&  5.21& 12.49&  0.04&0.41&0.35&0.65\\
 HD119802   &4763&4.0&--0.05&--0.04&+0.07&--0.06&  --4.70&  --9.90& --13.80&  8.10&  8.53&  0.09&0.03&0.98&0.02\\
 HD122064   &4937&4.5&+0.07&+0.01&+0.14&+0.09&  --2.70& --10.20& --26.60&  8.09&  8.57&  0.25&0.03&0.97&0.03\\
 HD125184   &5695&4.3&+0.31&--0.05&+0.01&+0.10&  36.10&   3.60& --43.50&  7.68& 10.49&  0.61&0.15&0.90&0.10\\
 HD126053   &5728&4.2&--0.32&+0.17&+0.12&--0.02&  22.20& --15.50& --39.30&  7.34&  9.12&  0.47&0.11&0.92&0.08\\
 HD131977   &4683&3.7&--0.24&+0.12&+0.18&+0.03&  48.20& --22.10& --32.70&  6.56&  9.76&  0.38&0.20&0.90&0.10\\
 HD135204   &5413&4.0&--0.16&+0.23&+0.13&+0.00& --85.40& --99.20& --15.00&  3.24&  9.26&  0.11&0.48&0.00&1.00\\
 HD135599   &5257&4.3&--0.12&+0.00&+0.05&--0.01&   9.70&   1.10& --13.70&  8.18&  9.35&  0.09&0.07&0.98&0.02\\
 HD137107   &6037&4.3&+0.00&+0.05&+0.06&--0.05&  14.90&  --5.50& --12.90&  7.87&  9.17&  0.08&0.08&0.98&0.02\\
 HD139323   &5204&4.6&+0.19&+0.05&+0.04&+0.12& --44.00& --64.30& --26.80&  4.82&  8.72&  0.26&0.29&0.30&0.70\\
 HD139341   &5242&4.6&+0.21&--0.02&+0.09&+0.11& --42.90& --67.00& --24.80&  4.70&  8.70&  0.23&0.30&0.25&0.75\\
 HD140538   &5675&4.5&+0.02&--0.02&+0.05&+0.05&  18.00&  --7.20&  10.50&  7.74&  9.22&  0.20&0.09&0.98&0.02\\
 HD141004   &5884&4.1&--0.02&+0.10&+0.08&+0.03& --49.10& --24.00& --39.80&  6.78&  9.19&  0.49&0.15&0.84&0.16\\
 HD144287   &5414&4.5&--0.15&+0.07&+0.12&+0.00& --96.60&  --9.30&  11.30&  6.45& 11.45&  0.25&0.28&0.89&0.11\\
 HD144579   &5294&4.1&--0.70&+0.37&+0.28&+0.05& --36.00& --58.60& --18.50&  5.11&  8.65&  0.15&0.26&0.57&0.43\\
\hline
\end{tabular}
\end{table}

\begin{table}[!hb]
\begin{tabular}{rrrrrrrrrrrrrrrr}
\hline
\hline
HD& \Teff & \logg &\FeH & \MgFe & \SiFe &   \NiFe & U & V& W & $\mathrm{R_{min}}$ & $\mathrm{R_{max}}$ &\Zmax &$\mathrm{ecc}$
& $\mathrm{Pr_1}$ & $\mathrm{Pr_2}$ \\
\hline
 HD145675   &5406&4.5&+0.32&+0.06&+0.09&+0.13&  23.70& --12.40& --16.00&  7.40&  9.22&  0.12&0.11&0.98&0.02\\
 HD146233   &5799&4.4&+0.01&+0.04&+0.05&+0.03&  25.90& --14.70& --23.10&  7.25&  9.22&  0.21&0.12&0.97&0.03\\
 HD149661   &5294&4.5&--0.04&+0.02&+0.09&+0.05&   1.00&  --0.40& --28.50&  8.36&  9.06&  0.29&0.04&0.97&0.03\\
 HD151541   &5368&4.2&--0.22&+0.05&+0.03&--0.06& --56.30& --10.50&  16.80&  7.17&  9.81&  0.31&0.16&0.96&0.04\\
 HD152391   &5495&4.3&--0.08&+0.06&+0.04&+0.01&  84.50&--110.40&   9.70&  2.81&  9.57&  0.20&0.55&0.00&1.00\\
 HD154345   &5503&4.3&--0.21&+0.07&+0.10&+0.00& --79.40&  --8.60& --36.10&  6.81& 10.80&  0.47&0.23&0.84&0.16\\
 HD154931   &5910&4.0&--0.10&+0.14&+0.08&+0.05&  11.80& --52.90& --19.10&  5.40&  8.54&  0.15&0.23&0.75&0.25\\
 HD157089   &5785&4.0&--0.56&+0.29&+0.24&--0.03&--166.90& --41.20& --10.30&  4.55& 13.70&  0.07&0.50&0.04&0.96\\
 HD158633   &5290&4.2&--0.49&+0.17&+0.15&+0.00&   1.50& --50.00&   5.50&  5.61&  8.53&  0.13&0.21&0.85&0.15\\
 HD159062   &5414&4.3&--0.40&+0.28&+0.22&+0.06& --26.40& --55.60& --60.30&  5.49&  8.59&  0.94&0.22&0.09&0.91\\
 HD159222   &5834&4.3&+0.06&+0.00&+0.08&+0.03& --30.80& --49.90&  --2.20&  5.56&  8.61&  0.04&0.22&0.83&0.17\\
 HD159482   &5620&4.1&--0.89&+0.37&+0.34&+0.02&--165.20& --62.80&  80.50&  4.19& 13.49&  2.78&0.53&0.00&0.98\\
 HD159909   &5749&4.1&+0.06&+0.06&+0.07&+0.01& --59.10& --55.00&  --6.90&  5.11&  9.00&  0.02&0.28&0.62&0.38\\
 HD160346   &4983&4.3&--0.10&--0.01&+0.09&+0.00&  19.40&   0.50&  11.60&  7.95&  9.63&  0.22&0.10&0.98&0.02\\
 HD161098   &5617&4.3&--0.27&+0.07&+0.09&--0.02&  --1.10& --40.90&  --7.10&  6.09&  8.49&  0.02&0.16&0.92&0.08\\
 HD164922   &5392&4.3&+0.04&+0.08&+0.11&+0.07&  60.10&   5.20& --48.00&  7.27& 11.60&  0.77&0.23&0.80&0.20\\
 HD165173   &5505&4.3&--0.05&+0.13&+0.10&+0.04&  --5.50&   2.90&  13.40&  8.46&  9.16&  0.24&0.04&0.98&0.02\\
 HD165341   &5314&4.3&--0.08&+0.03&+0.10&+0.03&   6.20& --18.80& --14.30&  7.39&  8.65&  0.09&0.08&0.97&0.03\\
 HD165401   &5877&4.3&--0.36&+0.26&+0.19&+0.05& --77.80& --89.50& --38.60&  3.64&  9.17&  0.48&0.43&0.00&1.00\\
 HD165476   &5845&4.1&--0.06&+0.01&+0.04&--0.04&  39.50& --21.10&  36.10&  6.80&  9.49&  0.67&0.17&0.90&0.10\\
 HD165670   &6178&4.0&--0.10&--0.03&+0.11&+0.01&  29.90&  --5.60& --14.00&  7.51&  9.63&  0.10&0.12&0.98&0.02\\
 HD165908   &5925&4.1&--0.60&+0.16&+0.10&--0.06&  --5.90&   0.70&   9.50&  8.48&  8.99&  0.19&0.03&0.98&0.02\\
 HD166620   &5035&4.0&--0.22&+0.13&+0.13&+0.00&  16.80& --31.30&   0.30&  6.54&  8.74&  0.07&0.14&0.96&0.04\\
 HD168009   &5826&4.1&--0.01&+0.03&+0.05&+0.02&  --4.60& --62.20& --22.60&  5.00&  8.49&  0.20&0.26&0.49&0.51\\
 HD173701   &5423&4.4&+0.18&+0.00&+0.17&+0.17&  --9.30& --46.70&  --3.20&  5.79&  8.49&  0.03&0.19&0.88&0.12\\
 HD176841   &5841&4.3&+0.23&+0.07&+0.12&+0.11&   9.30& --24.70&  12.30&  7.04&  8.66&  0.22&0.10&0.97&0.03\\
 HD182488   &5435&4.4&+0.07&+0.05&+0.10&+0.09& --20.90& --14.10&  --3.10&  7.70&  8.63&  0.03&0.06&0.98&0.02\\
 HD182736   &5430&3.7&--0.06&+0.04&+0.01&+0.05& --28.40& --39.10& --28.70&  6.17&  8.62&  0.28&0.17&0.85&0.15\\
\hline
\end{tabular}
\end{table}

\begin{table}[!hb]
\begin{tabular}{rrrrrrrrrrrrrrrr}
\hline
\hline
HD& \Teff & \logg &\FeH & \MgFe & \SiFe &   \NiFe & U & V& W & $\mathrm{R_{min}}$ & $\mathrm{R_{max}}$ &\Zmax &$\mathrm{ecc}$
& $\mathrm{Pr_1}$ & $\mathrm{Pr_2}$ \\
\hline
 HD183341   &5911&4.3&--0.01&+0.07&+0.09&+0.01& --23.00& --34.40&   1.50&  6.43&  8.54&  0.09&0.14&0.95&0.05\\
 HD184499   &5750&4.0&--0.64&+0.37&+0.34&+0.07& --64.90&--161.80&  58.60&  1.34&  8.96&  1.22&0.74&0.00&0.88\\
 HD184768   &5713&4.2&--0.07&+0.15&+0.15&+0.06&  25.20& --62.20& --27.80&  4.92&  8.69&  0.28&0.28&0.40&0.60\\
 HD185144   &5271&4.2&--0.33&+0.01&+0.06&--0.04&  31.70&  43.40& --19.10&  8.24& 13.90&  0.20&0.26&0.89&0.11\\
 HD186104   &5753&4.2&+0.05&+0.06&+0.07&+0.04& --53.20& --39.80& --20.30&  5.88&  9.02&  0.17&0.21&0.84&0.16\\
 HD186408   &5803&4.2&+0.09&+0.09&+0.06&+0.03&  17.90& --30.50&  --0.30&  6.58&  8.76&  0.06&0.14&0.96&0.04\\
 HD186427   &5752&4.2&+0.02&+0.09&+0.07&+0.05&  17.50& --30.60&  --1.80&  6.57&  8.75&  0.05&0.14&0.96&0.04\\
 HD187897   &5887&4.3&+0.08&+0.05&+0.08&+0.02& --41.10& --15.20&  --6.50&  7.25&  9.11&  0.01&0.11&0.97&0.03\\
 HD189087   &5341&4.4&--0.12&+0.02&+0.11&+0.00& --40.40& --15.00&   3.50&  7.29&  9.11&  0.11&0.11&0.97&0.03\\
 HD190360   &5606&4.4&+0.12&+0.06&+0.13&+0.10& --12.10& --44.90& --64.00&  6.15&  8.51&  1.02&0.16&0.17&0.83\\
 HD190404   &4963&3.9&--0.82&+0.39&+0.29&--0.03&  84.80& --46.00&  27.20&  5.13& 10.43&  0.53&0.34&0.49&0.51\\
 HD191533   &6167&3.8&--0.10&+0.03&+0.09&--0.03&  42.70& --26.20& --11.60&  6.43&  9.38&  0.07&0.19&0.95&0.05\\
 HD194598   &5890&4.0&--1.21&+0.28&+0.28&--0.11& --75.80&--276.40& --30.50&  0.99&  8.92&  0.46&0.80&0.00&0.00\\
 HD195005   &6075&4.2&--0.06&+0.12&+0.08&--0.07&  --6.10&  --8.20&  --6.70&  8.22&  8.50&  0.01&0.02&0.98&0.02\\
 HD195104   &6103&4.3&--0.19&+0.16&+0.07&--0.09& --14.90&  --6.60&   3.00&  8.24&  8.62&  0.10&0.02&0.98&0.02\\
 HD197076   &5821&4.3&--0.17&+0.07&+0.09&+0.01& --43.20& --15.10&  16.30&  7.25&  9.21&  0.29&0.12&0.97&0.03\\
 HD199960   &5878&4.2&+0.23&+0.02&+0.10&+0.11&  --7.40& --24.50&  --2.60&  7.10&  8.48&  0.04&0.09&0.97&0.03\\
 HD201889   &5600&4.1&--0.85&+0.40&+0.33&--0.02&--129.40& --82.30& --37.20&  3.59& 10.71&  0.50&0.50&0.00&1.00\\
 HD201891   &5850&4.4&--0.96&+0.29&+0.32&--0.03&  92.00&--115.10& --58.60&  2.71&  9.82&  1.03&0.57&0.00&0.99\\
 HD202108   &5712&4.2&--0.21&+0.06&+0.05&--0.05&  --7.50&   6.50&  10.40&  8.49&  9.47&  0.20&0.05&0.98&0.02\\
 HD203235   &6071&4.1&+0.05&+0.10&+0.11&--0.01& --46.30& --18.70& --18.40&  6.97&  9.19&  0.15&0.14&0.96&0.04\\
 HD204521   &5809&4.6&--0.66&+0.27&+0.20&+0.04&  15.40& --73.20& --19.00&  4.46&  8.61&  0.15&0.32&0.21&0.79\\
 HD205702   &6020&4.2&+0.01&+0.10&+0.08&+0.06&  29.60& --29.50&   6.90&  6.49&  8.99&  0.16&0.16&0.96&0.04\\
 HD208906   &5965&4.2&--0.80&+0.23&+0.25&+0.04&  73.20&  --2.10& --11.00&  6.79& 11.55&  0.07&0.26&0.95&0.05\\
 HD210667   &5461&4.5&+0.15&--0.10&+0.06&+0.07&  13.70& --25.10& --16.50&  6.93&  8.73&  0.12&0.11&0.96&0.04\\
 HD210752   &6014&4.6&--0.53&+0.07&+0.11&+0.02& --10.50&  27.70& --63.10&  8.49& 12.01&  1.22&0.17&0.48&0.52\\
 HD211472   &5319&4.4&--0.04&+0.00&+0.05&--0.01& --20.00& --12.10&  --6.00&  7.84&  8.65&  0.00&0.05&0.98&0.02\\
 HD215065   &5726&4.0&--0.43&+0.31&+0.18&+0.03& --31.50& --66.10&  26.80&  4.83&  8.61&  0.46&0.28&0.29&0.71\\
\hline
\end{tabular}
\end{table}

\begin{table}[!hb]
\begin{tabular}{rrrrrrrrrrrrrrrr}
\hline
\hline
HD& \Teff & \logg &\FeH & \MgFe & \SiFe &   \NiFe & U & V& W & $\mathrm{R_{min}}$ & $\mathrm{R_{max}}$ &\Zmax &$\mathrm{ecc}$
& $\mathrm{Pr_1}$ & $\mathrm{Pr_2}$ \\
\hline
 HD215704   &5418&4.2&+0.07&+0.04&+0.06&+0.05& --17.00& --58.90&  --9.00&  5.16&  8.52&  0.03&0.25&0.67&0.33\\
 HD217014   &5778&4.2&+0.14&+0.06&+0.08&+0.09& --15.20& --29.70&  15.60&  6.81&  8.52&  0.27&0.11&0.95&0.05\\
 HD218209   &5705&4.5&--0.43&+0.19&+0.18&+0.04& --72.80& --47.00& --14.50&  5.36&  9.45&  0.10&0.28&0.68&0.32\\
 HD219134   &4900&4.2&+0.05&--0.02&+0.09&+0.04& --52.30& --39.70& --14.10&  5.91&  9.02&  0.09&0.21&0.87&0.13\\
 HD219396   &5733&4.0&--0.10&+0.16&+0.10&+0.03&   3.40& --53.60&  --4.20&  5.43&  8.55&  0.03&0.22&0.80&0.20\\
 HD224930   &5300&4.1&--0.91&+0.42&+0.29&+0.01&  --8.70& --71.30& --32.40&  4.60&  8.50&  0.34&0.30&0.16&0.84\\
\hline
\end{tabular}
\end{table}
\endlandscape

\end{document}